\documentclass{aa}  

\usepackage{longtable}
\usepackage{graphicx}
\usepackage{txfonts}
\usepackage{longtable,lscape}
\usepackage{booktabs}

\def\sun{\hbox{$\odot$}}
\def\R23{\mbox{$\rm R_{23}$}}

\def\kmsmpc{km s$^{-1}$ Mpc$^{-1}$}

\def\msun{M$_{\odot}$}

\def\Hb{\mbox{${\rm H}{\beta}$}}
\def\Ha{\mbox{${\rm H}{\alpha}$}}
\def\OIIIa{\mbox{${\rm [O\,III]\,}{\lambda\,5007}$}}

\def\OII{\mbox{${\rm [O\,II]\,}{\lambda\,3727}$}}

\def\NII{\mbox{${\rm [N\,II]\,}{\lambda\,6584}$}}

\begin{document}

\title{CLASH-VLT: Strangulation of cluster galaxies in  MACS\,J0416.1-2403 as seen from their chemical enrichment
\thanks{Based on observations obtained at the European Southern Observatory (ESO) Very Large  Telescope (VLT), Paranal, Chile; ESO large program 186.A-0798}}


\author{C.~Maier\inst{1}
\and U.\,~Kuchner\inst{1}
\and B.\,L.~Ziegler\inst{1}
\and M.\,~Verdugo\inst{1}
\and I.\,~Balestra\inst{2}
\and M.\,~Girardi\inst{2,3}
\and A.\,~Mercurio\inst{4}
\and P.\,~Rosati\inst{5}
\and A.\,~Fritz\inst{6}
\and C.\,~Grillo\inst{7}
\and M.\,~Nonino\inst{2}
\and B.\,~Sartoris\inst{2,3}
}

\institute{University of Vienna, Department of Astrophysics, Tuerkenschanzstrasse 17, 1180 Vienna, Austria\\
\email{christian.maier@univie.ac.at}
\and INAF - Osservatorio Astronomico di Trieste, via G. B. Tiepolo 11, I-34133, Trieste, Italy
\and Dipartimento di Fisica, Universit\`a degli Studi di Trieste, via Tiepolo 11, I-34143 Trieste, Italy
\and INAF - Osservatorio Astronomico di Capodimonte, Via Moiariello 16, I-80131 Napoli, Italy
\and Dipartimento di Fisica e Scienze della Terra, Universit\`a di Ferrara, Via Saragat 1, I-44122 Ferrara, Italy
\and INAF - Istituto di Astrofisica Spaziale e Fisica cosmica (IASF) Milano, via Bassini 15, I-20133 Milano, Italy
\and Dark Cosmology Centre, Niels Bohr Institute, University of Copenhagen, Juliane Maries Vej 30, DK-2100 Copenhagen, Denmark
}

\titlerunning{MZR in cluster galaxies at $z \sim 0.4$}
\authorrunning{C. Maier et al.}

\date{Received ; accepted}

\abstract 
{}
{Environmental effects 
gain importance as large scale structures in the Universe develop with time 
and have become the dominant mechanism for quenching galaxies of intermediate and low stellar masses at lower redshifts. Therefore, clusters of galaxies at $z<0.5$ are the sites where environmental effects are expected to be more pronounced and more easily  observed with present-day large telescopes.}
{We explore the Frontier Fields cluster MACS\,J0416.1-2403 at $z=0.3972$ with VIMOS/VLT spectroscopy from the CLASH-VLT survey covering a region 
 that corresponds to almost three virial radii. We measure fluxes
of 
\Hb, \OIIIa, \Ha, and \NII\, emission lines  of cluster members enabling us to unambiguously derive O/H gas metallicities, and also star formation rates from extinction-corrected \Ha\, fluxes. We compare our cluster galaxy sample with a field sample at $z \sim 0.4$ drawn from zCOSMOS.}
{The 76 galaxies of our cluster sample follow the star-forming metallicity sequence in a diagnostic diagram disentangling ionizing sources.
For intermediate masses we find a similar distribution of 
cluster and field galaxies in the  mass vs. metallicity  and mass vs. sSFR  diagrams. 
An in-depth investigation furthermore reveals that bulge-dominated cluster galaxies have on average lower sSFRs and higher O/Hs than  their disk-dominated counterparts.
We use the location of galaxies in the projected velocity vs. position phase-space to separate our cluster sample into a region of objects accreted longer  ago and a region of recently accreted and infalling galaxies.
We find a higher fraction  of accreted metal-rich galaxies  (63\%) compared to the fraction of 28\% of metal-rich galaxies in the infalling regions.
Intermediate-mass galaxies ($9.2<\rm{log(M/M_{\odot})}<10.2$)  falling into the cluster for the first time are found to be in agreement with predictions of the fundamental metallicity relation. In contrast, for already accreted star-forming galaxies of similar masses, we find on average metallicities higher than predicted by the models.
This trend is intensified for accreted cluster galaxies of the lowest mass bin 
($\rm{log(M/M_{\odot})}<9.2$), that display metallicities two to three times higher than predicted by models with primordial gas inflow. Environmental effects therefore strongly influence gas regulations and control gas metallicities of $\rm{log(M/M_{\odot})}<10.2$ cluster galaxies.
We also investigate chemical evolutionary paths of model galaxies with and without inflow of gas showing that strangulation is needed to explain the higher metallicities of accreted cluster galaxies.
Our results favor  a strangulation scenario in which gas inflow stops for galaxies with $\rm{log(M/M_{\odot})}<10.2$ when accreted by the cluster.%
}{}


\maketitle



\setcounter{section}{0}
\section{Introduction}

~~~Gas-phase metallicites in galaxies are regulated by the complex interplay between star formation, accretion of metal-poor gas and galactic outflows of processed gas in galaxies, representing their current state of chemical enrichment.
Comprehensive studies of large samples in the local Universe have established a tight relation between stellar mass and metallicity 
showing that the brighter and hence more massive a galaxy is, the greater is its gas metallicity as measured by the O/H abundance \citep[e.g., using SDSS,][]{trem04,keweli08}.
One explanation for the origin of this 
relation has been summarized by \citet[][Li13 in the following]{lilly13}, and is known as the bathtub model. The star formation rate (SFR) is thereby closely linked 
to the mass of gas in the galaxy  modulated
by infalling gas and outflows driven by supernovae winds. In this scenario  a relation between metallicity, SFR, and stellar mass can be derived.
Several authors have studied the role of the SFR  and of the specific SFR (sSFR$=$SFR/M$_{*}$) in this context and have presented links between SFR and metallicity \citep[e.g.,][]{elli08,lopez10}, finally claiming an epoch-independent fundamental metallicity relation (FMR) between metallicity, mass, and SFR, Z(M$_{*}$,SFR), expected to be applicable at all redshifts  \citep[][]{mannu10}.
Recenty, \citet{salim14} reconciled conflicting results regarding this SFR dependence showing that the metallicity is anti-correlated with sSFR for SDSS local galaxies, regardless of the metallicity and SFR indicators used.

The evolution of galaxies is governed internally by their stellar mass and externally influenced by their environment \citep[e.g.,][]{peng10,peng12} made visible through well-established changes in the observable properties of cluster galaxies that affect the star formation histories (SFHs), gas and stellar components.
Therefore, examining the interrelationships between stellar mass, environment, metallicity, and stellar structure, and their evolution with time will reveal the physical processes that control the evolution of galaxies. %
Since environmental effects get stronger at $z<0.5$ as large-scale structure in the Universe develops \citep[see, e.g., Fig.\,15 in][]{peng10}, clusters of galaxies at $z<0.5$ are the sites where influences of the environment should be  detected more easily with the available present-day large telescopes.
It is well established that specific cluster phenomena influence star formation  and the exchange of gas, for example, by supressing gas inflow. Star formation and gas inflow influence gas metallicities; therefore, it is expected that gas metallicities depend on environment.

According to previous works in the \textit{local} Universe, using SDSS galaxies, little effect of the environment on the mass-metallicity relation (MZR) was found.
Defining environment by local densities, \citet{mouhcine07} reported variation in metallicity, at a given mass, ranging from 0.02\,dex for massive to 0.08\,dex for galaxies with $\rm{logM/M_{\odot}} < 9.5$ occurring over a contrast factor of 100. 
In their environmental studies of central and satellite SDSS galaxies, \citet{pasq12} and \citet{pengmaio14} both reported an average metallicity of satellites  higher than that of centrals, especially for low stellar mass galaxies; this difference in metallicity was found to gradually
disappear towards higher stellar masses.
Another investigation of 1318 galaxies in cluster environments found an 
increase of 0.04\,dex in oxygen abundances for galaxies in clusters compared to the field \citep{elli09}.
This was confirmed numerically
 by \citet{dave11}, who used cosmological hydrodynamic simulations to investigate environmental effects on the MZR, reporting a small enhancement in the mean MZR of galaxies in high-density regions of $\sim 0.05$\,dex.
On the other hand, \citet{cooper08} used the SDSS sample and 
claimed that there is a stronger 
relationship between metallicity and environment (largely driven by galaxies in high-density regions such as groups and clusters), such that more metal-rich galaxies favor regions of higher overdensity.  


The current  state and gas content of galaxies is furthermore expected to be related to their morphologies. The importance of structural parameters such as compactness and surface mass densities in setting the gas properties has been recognized before \citep{saint12}. Galaxies with higher surface mass densities have transformed more of their gas into stars over the course of their SFH, thereby enhancing the overall gas-phase metallicity \citep{kaufm03b}.
The dependence of the MZR on structural parameters leaves its imprint on the scatter of the relation. \citet{trem04} related higher stellar mass surface densities to higher metallicities at fixed stellar mass. Additionally, \citet{elli08} found that at fixed stellar masses, galaxies with larger sizes (i.e., with lower stellar mass surface densities) have lower metallicities.
On the theroretical side,  \citet{calura09} studied the MZR by means of chemical evolution models of different morphological types.
 Their results indicate that the  observed MZR is reproduced by considering different morphological mixes at different redshifts.
The varying strength of a central bulge governs the classification scheme introduced by \citet{hubble26}.
While connections between the presence of a bulge  and quenching of star formation have been reported \citep[e.g.,][suggesting that the main driver for gas depletion in galaxies is bulge mass]{saint12}, the physical mechanisms responsible for this star formation shutoff and the enhancement of measured gas metallicities is not yet fully understood.
Connecting the gas metallicities with mass, SFRs, and morphologies as a function of environment and redshift may therefore lead to increased knowledge of the temporal evolution of the chemical properties of the stellar populations and allow a more qualitative look into the nature of the galaxies building the MZR.


We can now extend these investigations to intermediate redshifts, measuring O/H abundances in a very massive cluster  at $z \sim 0.4$, 
and contrasting them to comparable measurements of zCOSMOS field galaxies at similar redshifts. We do this by using the two connected relations, the MZR and the mass$-$(s)SFR relation, presenting evolutionary effects between $z \sim 0.4$ and the local Universe.
To estimate the chemical abundances, a number of 
diagnostics have been developed based on strong emission lines (ELs), 
 ${\rm [O\,II]\,}{\lambda\,3727}$, H$\beta$, ${\rm [O\,III]\,}{\lambda\,5007}$, H$\alpha$, and [NII]${\lambda\,6584}$.
At higher redshifts, these ELs move to the near-infrared, with H$\alpha$ and [NII]${\lambda\,6584}$ already shifting beyond the optical above redshifts of $\sim 0.5$, requiring near-infrared spectroscopy to observe these ELs \citep[e.g.,][]{maier04,maier05,maier06,maier14,maier15}. Thus, the epoch we probe is the highest redshift at which all these ELs can be observed 
with the same optical spectrograph 
without the complication of obtaining a good relative calibration between spectra from optical and near-infrared spectrographs.


  The paper is structured as follows. In Sect. 2 we
present the selection of the cluster EL galaxies at $z \sim 0.4$ and their VIMOS spectroscopy.
We also discuss the comparison field samples of zCOSMOS and SDSS galaxies.
In Sect.\,3 we investigate the active galactic nucleus (AGN) contribution
and present the derivation of SFRs, metallicities, morphological parameters, and stellar masses of the 76 cluster galaxies we observed.
In Sect.\,4 we present  the
MZR and mass$-$sSFR relation at $z\sim 0.4$, and investigate their dependence on environment and 
morphological parameters. 
In Sect.\,5  we discuss how the cluster environment affects the chemical enrichment. We investigate how the SFR impacts the MZR in the cluster environment, and how this compares with  predictions of the FMR.  Several pieces of evidence for a strangulation scenario for intermediate- and low-mass cluster galaxies are discussed.
Finally in Sect.\,6 we summarize our conclusions.
A {\sl concordance}-cosmology with $\rm{H}_{0}=70$ \kmsmpc,
$\Omega_{0}=0.25$, $\Omega_{\Lambda}=0.75$ is used throughout this
paper.  
We assume a Salpeter \citep{salp55} initial mass function (IMF) for all derived stellar masses and SFRs  and correct existing measurements used in this paper to a Salpeter IMF. 
We note that {\sl
  metallicity} and {\sl abundance} will be taken to denote {\it oxygen abundance}, O/H, throughout this paper, unless otherwise specified.
In addition, we use dex throughout to denote the anti-logarithm,
i.e., 0.3\,dex is a factor of two.

\section{The data: cluster and field galaxies at $z \sim 0.4$}

\subsection{The  MAC0416 cluster}

~~~MACS\,J0416.1-2403 (hereafter MAC0416) is a massive, X-ray-luminous \citep{ebeling01} galaxy cluster at $z=0.3972$.
MAC0416 is one of 25 clusters in the HST multicycle treasury program ``Cluster Lensing And Supernova survey with Hubble'' \citep[CLASH,][]{postman12}. 
The mass profile measured from the galaxy dynamics \citep[][hereafter B16]{balest16} reveals a complex dynamical state of this cluster that can be explained by a  pre-collisional phase scenario.
The inner regions of the cluster are nevertheless represented best by a simple 
isothermal sphere (SIS) model, as described in B16.

CLASH-VLT \citep{rosati14}, the spectroscopic follow-up of 13 CLASH clusters at $0.3<z<0.6$, provides spectra over an area of $25 \times 25$ square arcminutes, using the Low Resolution Blue (hereafter LB)  and Medium Resolution (hereafter MR) grisms of VIMOS.
The details of the LB and MR VIMOS observations and redshift determinations of MAC0416 are given in  B16. The observations and data reduction of the MR pointings used for our SFR and metallicity measurements are described in Sect.\,\ref{sec:obsredu}.
High quality SUBARU Suprime-Cam BRz imaging covers the VIMOS field that is  $25 \times 25$ square arcminutes in area \citep[][]{umetsu14}.
For our metallicity study we selected this particular cluster  from the CLASH sample   owing to its ideal redshift ($z \sim 0.4$), which places all five strong ELs in a region not affected by night skylines, allowing measurements of all ELs needed in one spectrum  using the MR grism of VIMOS.

The identification of cluster members  was described in B16, adopting 
as the center of the cluster the position of one of the two brightest cluster galaxies (BCGs), namely the NE-BCG position, which coincides with the peak of the X-ray emission \citep[see][]{ogrean15}.
\citet{balest16} obtained a velocity dispersion $\sigma \sim 1000$\,km/s,  virial radius $R_{200} \sim 1.8$\,Mpc, and 
a virial mass of $M_{200} \sim 1\times 10^{15}$\,M$_\odot$ for MAC0416, where the ``200'' subscript refers to the radius where
the average density is 200 times the mean density of the Universe at the cluster redshift.
We consider as cluster members all galaxies within a larger range of $\pm 3\sigma$  of the velocity dispersion because we want to study environmental effects  by dividing the sample into ``infalling'' and ``accreted longer  ago'' cluster galaxies, as described  in Sect.\,\ref{sec:discussion}.


\subsubsection{Selection of the cluster galaxies for our metallicity study}

~~~The target selection for the LB VIMOS pointings was based on color-color diagrams (as described in B16).
%
The spectroscopic redshifts derived from the \OII\, ELs as delivered by the VIMOS LB grism observations allowed us to select an appropriate  target sample, with ELs free of strong telluric emission, for the MR grism observations that provide  higher resolution and larger (including redder) wavelength coverage than the LB spectra. The target sample contained 165 blue cluster member galaxies with the four ELs \Hb, \OIIIa, \Ha, and \NII, which are necessary for the derivation of reliable metallicities, free of contamination by night skylines.
Eighty-two galaxies in this sample were observed using the MR grism, 60 of which   have a signal-to-noise ratio (S/N) higher than two in the weakest EL detected (\Hb, \OIIIa, or \NII).
Additionally, some galaxies without LB spectra were selected based on color-color criteria and were observed with the MR grism  without being part of the parent sample of 165 blue cluster members. Sixteen of these galaxies turned out to have five ELs with high enough  S/N for metallicity studies resulting in a total sample of 76 cluster galaxies with five ELs detected (sample $Zgals$).

\begin{figure*}
\includegraphics[width=\textwidth,angle=0,clip=true]{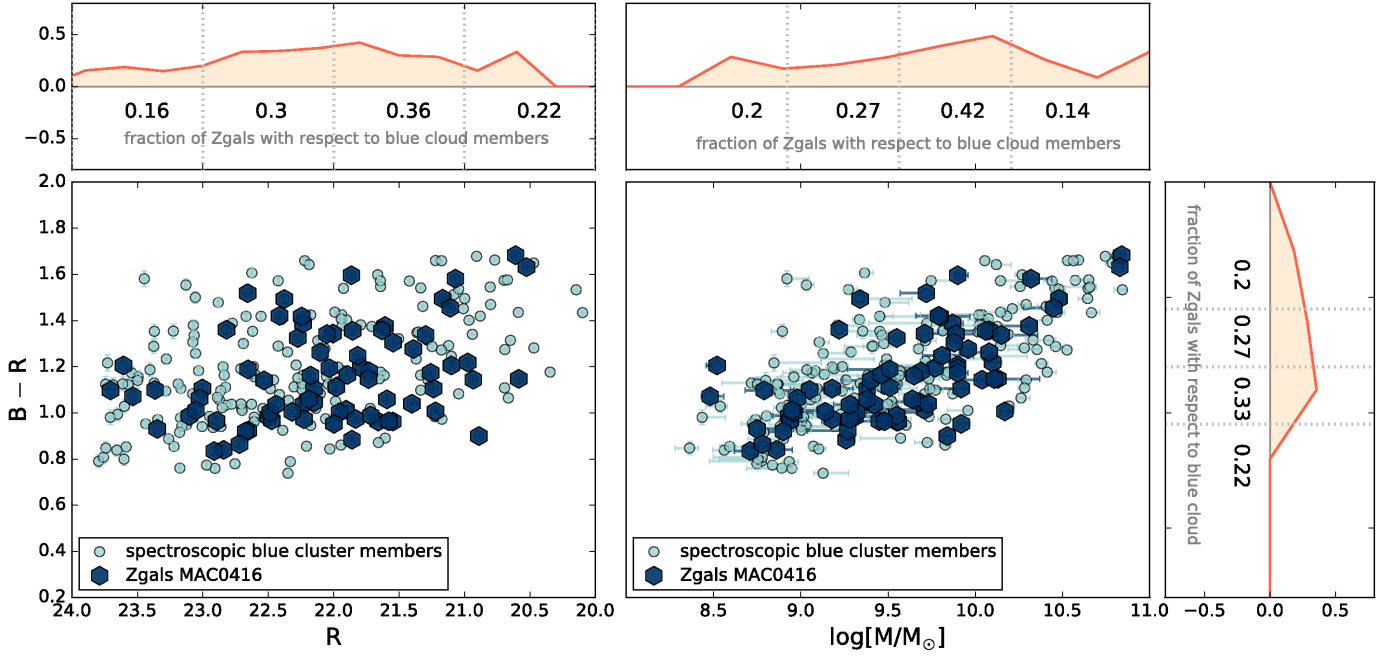}
\caption
{
\label{Selec0416} 
\footnotesize 
Color-luminosity (left) and color-mass (right) diagrams for MAC0416 cluster galaxies at $z \sim 0.4$  with spectroscopic redshifts.
Cyan data points are the spectroscopic confirmed MAC0416 blue cluster members, representative of the star-forming blue cloud galaxies, while the 76 galaxies used for this metallicity study  ($Zgals$ sample) are shown in blue. The red lines and numbers indicate the fraction of $Zgals$ objects (blue symbols) with respect to the parent sample (cyan points). We achieve a uniform coverage at intermediate luminosities/masses. The lowest and highest luminosity/mass bins are under-represented by our sample.
}
\end{figure*}

Fig.\,\ref{Selec0416} demonstrates the uniform coverage of our sample $Zgals$ with respect to the parent sample of all
spectroscopic blue cloud galaxies in both luminosities  and masses.
The figure shows the color-magnitude and color-mass diagrams of the spectroscopic member galaxies of the blue
cloud (cyan filled circles) and the 76 $Zgals$ objects used for our metallicity study (blue symbols). The fractions indicate that our $Zgals$ sample is a good representation of the overall 
blue (star-forming) population of  cluster galaxies, especially at intermediate masses. 
Table\,1 lists the percentages of all
spectroscopic members representative for the blue cloud (first row) in four luminosity, mass, and color bins,  the percentage
of the $Zgals$ objects in the same bins (second row), and their fractions (third row) as visualized in Fig.\,\ref{Selec0416}.


\subsubsection{MR VIMOS observations and data reduction}
\label{sec:obsredu}

~~~The main target galaxies were observed 
in four VIMOS pointings with the MR grism and GG475 filter  with exposure times of 60 min each, and masks designed with 1 arcsec slits.
The covered wavelength range was 480-1000\,nm with a resolution $R=580$, thereby including ELs from \OII\, to \NII\, of $z \sim 0.4$ galaxies.

Data reduction of the VIMOS spectra of MAC0416 galaxies was carried out using the Vimos Interactive
Pipeline Graphical Interface \citep[VIPGI;][]{scod05} software
package. VIPGI uses automated  procedures for bias subtraction, flux and wavelength calibration of the spectra, identification of objects in the slit profile, and extraction of the one-dimensional spectra. Wavelength calibration uses a HeNeAr arc lamp exposure obtained immediately following the science exposures. The standard star LTT-1788 was used for flux calibration.
The determination of redshifts of target galaxies is unambiguous owing to the five ELs observed with the MR grism of VIMOS.


\subsection{Comparison sample: zCOSMOS field galaxies at $z\sim 0.4$}
\label{sec:zCOSMOS}

~~~As our
comparison sample of field galaxies, we selected 93 zCOSMOS-bright \citep{lilly07,lilly09} objects at similar redshifts to the cluster galaxies, namely $0.373<z<0.413$, ensuring that all ELs are observed free of strong OH night skylines.
 These objects have four ELs measured with VIMOS -- \Hb, \OIIIa, \Ha, and \NII\ -- and stellar masses derived from the SED fitting based on 30 COSMOS bands \citep{ilbert09}. 
For details of the measurements of zCOSMOS EL fluxes we refer to \citet{maier09}, \citet{perez13}, and \citet{maier15}.


\subsection{Local comparison sample: SDSS}
\label{sec:SDSS}

~~~We  selected a local comparison sample of SDSS galaxies from the DR4
release, Garching repository\footnote{http://www.mpa-garching.mpg.de/SDSS/DR4/},  in the redshift
range $0.04<z<0.08$. The lower redshift limit of $z = 0.04$ was chosen
following \citet{kew05} to reduce the systematic and random errors in the SFRs that arise from
aperture effects due to the 3 arcsec size of the SDSS fibers. Duplicate objects and SDSS galaxies
on problematic or special plates were excluded from the SDSS sample; since this exclusion was based solely on position on the sky, it  should not be related to any galaxy properties.

We excluded AGNs from the SDSS sample using the 
\OIIIa/ \Hb\, vs. \NII/\Ha\,  diagnostic diagram, excluding objects
that satisfy  Eq. (1) of \citet{kaufm03}.
The evolution of the sSFR shows an average decrease of a factor of about two
from $z \sim 0.4$ until the epoch of the SDSS sample, at redshifts $0.04<z<0.08$ 
(see also Sect.\,\ref{SSFRMsel}).
Therefore, a lower limit in SFR of 
$0.3M_{\odot}/yr$ 
for our $Zgals$ MAC0416 sample at $z \sim 0.4$ corresponds to a lower limit of $0.15M_{\odot}/yr$ for the SDSS sample.
To ensure a consistent comparison of the physical properties of the SDSS and $Zgals$ samples, we applied an additional $S/N>10$ cut in \Ha\, for SDSS galaxies as was applied for the $Zgals$ sample, and only SDSS galaxies with SFRs larger than $0.15M_{\odot}/yr$ were used.
Therefore, it should be noted that this sample is slightly different from other local comparison SDSS samples often used in the literature such as the samples of \citet{trem04} or \citet{mannu10}.


\section{ Measurements}

\subsection{EL fluxes}
\label{sec:elflux}

~~~The EL fluxes of MAC0416 cluster member galaxies
were measured interactively
using the package \emph{splot} in IRAF.  The flux errors were usually dominated by systematic uncertainties in establishing the local continuum, which was
conservatively estimated by exploring rather extreme possibilities.
Following the recommendation of \citet{kobul99} for the statistical correction of \Hb\, EL flux measurements from global galaxy spectra, we assumed an average underlying stellar absorption in \Hb\, of $3\pm2$\AA\, and corrected the 
equivalent width (EW) of this
line by this amount, thereby increasing the \Hb\, line flux. This correction is often used in the literature  for data with similar spectral resolution to that used in our sample \citep[e.g.,][]{lilly03,maier05}, and the range of 1\AA\, to 5\AA\, for the EW in \Hb\, absorption corresponds to the values found for the entire range of typical ages of the stellar population \citep[see, e.g.,][]{kobul99,maier15}. The uncertainty of 2\AA\, acts as an additional error term for lower S/N in \Hb. We use this uncertainty and account for it in the total error budget  when computing the error bars for the corrected \Hb\, EL fluxes and derived quantities (see Table\,\ref{MeasurementsMAC0416}).
We note that all  \Ha\, ELs of $Zgals$ cluster galaxies have a $S/N>10$, and 70 out of 76 galaxies (92\% of the sample) have a $S/N>3$ in \NII.

\subsection{Star formation or AGNs?}
\label{sec:BPT}

~~~We use the BPT \citep[Baldwin, Phillips \&Terlevich;][]{bald81} diagram shown in Fig.\,\ref{fig:BPT},
to establish whether the source of gas ionization is of stellar origin
or is instead associated with AGN activity.
The MAC0416 $Zgals$ cluster galaxies and also all zCOSMOS 
field galaxies at $z \sim 0.4$
lie under and to the left of the theoretical curve (solid)  of \citet{kewley01} and the empirical curve (dashed)  of \citet{kaufm03}, 
which separate star-forming galaxies from AGNs.
This indicates that in all of
them the dominant source of ionization of the gas is recent star
formation.
Moreover, most galaxies lie on the  star-forming abundance sequence with rough boundaries indicated by the thick dashed lines.
 zCOSMOS field galaxies, which all have $\rm{log(M/M_{\odot})}>9.2$, and $Zgals$ cluster galaxies of similar masses (blue symbols) lie in a similar, lower part of the BPT diagram. Cluster galaxies with lower masses $\rm{log(M/M_{\odot})}<9.2$ (yellow symbols) have typically higher \OIIIa/\Hb\, ratios, which correspond to lower metallicities \citep[cf. Fig. A.1 in][]{maier15}.


\subsection{Oxygen abundances}
\label{Oxabund}

~~~Equation\,4 in \citet{kewley13a}, $12+\rm{log}(O/H)=8.97-0.32 \times \rm{log}(([OIII]/\Hb)/([NII]/\Ha))$, corresponds to the \citet{petpag04} O3N2 calibration for the \citet{kewdop02} metallicity scale. The only difference from the \citet{petpag04} O3N2 calibration is that the first (normalization) factor is 8.97 instead of 8.73.
 We use this equation to consistently compute metallicities for the $Zgals$ MAC0416 cluster objects,  the zCOSMOS field galaxies at $z \sim 0.4$, and for the SDSS sample.

There are several reasons for this choice of metallicity calibration. First, the zCOSMOS sample at 
$z \sim 0.4$
does not have  [OII] measured, but only the other four ELs at redder wavelengths.
Second, 
it is more straightforward to compute O3N2 metallicities than to use  a 
$\chi^{2}$  method, which is based on the comparison of five measured ELs with theoretical predictions of the fluxes for given metallicities and extinction values. The latter method can be affected by gradients and degeneracies of the theoretical models \citep[see discussion in][]{maier05}. 
Third, using O3N2 the comparison with other literature data is easier, because the $\chi^{2}$ approach of simultaneously fitting all  five strong ELs and using theoretical models to generate a probability distribution of metallicities and statistically estimate the oxygen abundances is very seldom used in the literature.
Fourth, as discussed above in Sect.\,\ref{sec:BPT}, most galaxies in this study follow a star-forming abundance sequence in the BPT diagram, indicating that some line ratios are related to each other,  which the $\chi^{2}$ method does not take into account.


\subsection{Stellar masses}
\label{sec:masses}

~~~Stellar masses of the MAC0416  galaxies were calculated using the code 
\emph{Lephare} of \citet{arnilb11}, which fits 
stellar population synthesis
models \citep{bruzcharl03} to the available photometry 
(SUBARU B, R and z-bands). This code is a simple
$\chi^{2}$ minimization algorithm that finds 
the best match of templates for the given data. The limited number of filters results in a reduction of used templates. 
In particular, we do not fit templates with stellar ages of less than 1\,Gyr and larger than 11\,Gyrs. This is a valid assumption since 
at $z=0.4$ the Universe is $\sim$4 Gyrs younger than today, and a considerable fraction of the stellar mass assembly is already completed. In addition, we  kept the redshift fixed and also limited the number of extinction $E_{(B-V)}$ values (0.1, 0.3, 0.5) to avoid over-fitting.
We are confident in the robustness of the calculated masses since 
the $B-R$ color encompasses the redshifted 4000\AA\, break and thus is sensitive to galaxy mass-to-light ratios \citep{kaufm03b}.
We list in Table\,\ref{MeasurementsMAC0416} the uncertainties on the stellar masses corresponding to the 68\% confidence level on the mass, calculated by scanning  the $\chi^{2}$  levels, while the redshift for each galaxy is kept fixed, but the other parameters (dust extinction, SFR, age, metallicity) are allowed to vary.

To test the presence of any bias due to the limited broad-band photometry, we produced synthetic B, Rc, and z-band magnitudes from \citet{bruzcharl03} models with a range of stellar masses, ages, metallicities, star formation histories, and extinction values using the \emph{EzGal} tool \citep{mancgonz12}. These magnitudes were fed to \emph{Lephare} and then the output was compared to the input values. Overall stellar masses are more resistant than  other parameters, and show a scatter of about 0.1\,dex. This additional error is generally smaller than the typical uncertainties reported in Table\,\ref{MeasurementsMAC0416}.

As described in \citet{maier15}, the stellar masses for zCOSMOS galaxies were derived by fiting photometric data points with the \citet{bruzcharl03} synthetic stellar population models, and picking the best-fit parameters by minimizing the $\chi^{2}$ between observed and model fluxes. 
The stellar masses were obtained by integrating the SFR over the galaxy age, and correcting for the mass loss over the course of stellar evolution. Thus, the stellar masses of zCOSMOS galaxies are comparable to the stellar masses of MAC0416 galaxies, which are also based on SED fitting using \citet{bruzcharl03} models.
We converted the masses from a \citet{chabrier03} IMF  to a \citet{salp55} IMF (as used in this paper) by multiplying the \citet{chabrier03} IMF masses by 1.7. This factor of 1.7 was found by \citet{pozzetti07} to be a systematic median offset, with a very small dispersion, in the masses derived with the two different IMFs; it is  
rather constant for a wide range of SFHs.
Regarding the consistency of SDSS stellar masses with stellar masses derived from SED fitting, 
we note that \citet{moust13}  showed the comparison between the two methods of estimating stellar masses
indicating that they are consistent within the uncertainties.


\subsection{Morphologies}
\label{sec:morph}
~~~The decomposition of galaxies into their main stellar components is done by analyzing the light distribution with the sum of two S\'ersic profiles with different indices (e.g., a DeVaucouleurs bulge and an exponential disk). 
To quantify stellar structures of MAC0416 galaxies, we measure parameters such as S\'ersic indices, sizes, and luminosities  for the entire galaxy, and  for bulges and disks separately for all available bands (SUBARU B, R, and z) with the new galaxy profile model fitting software Galapagos-2 developed within the MegaMorph project \citep{haus13}. It performs  two-dimensional modeling of the galaxy flux assuming predefined light distributions for all bands available simultaneously. Galapagos-2 is constructed in a way to efficiently handle modern large multiwavelength imaging surveys and is thus ideally suited to analyze the CLASH data. It utilizes GALFIT-M \citep{bamf11}, a recently developed multiwavelength version of the two-dimensional fitting algorithm GALFIT \citep{pengCY10}. This novel approach was successfully tested on both ground-based and simulated data \citep{haus13,vika14} and leads to an increase in the stability and accuracy of measured parameters down to fainter limits in comparison to the widely used single-band fits. This is especially important for the measurements of ground-based SUBARU imaging for this study. 

To ensure reliable morphologies, we compare our SUBARU-based classification step by step with HST high-resolution CLASH data. 
However, we note  that none of the galaxies in our $Zgals$ sample is in the central part of the cluster covered by HST.
We therefore analyze an independent sample of about 100 overlapping galaxies of the central cluster region and find that 10\% (15\%) of the galaxies are misclassified using bulge-to-disk decompositions (single S\'ersic fits). 
Whether we classify based on a single S\'ersic fit or bulge-to-disk decomposition only changes the outcome in 5\% of the cases (based on measurements of 820 galaxies in CLASH clusters covered with SUBARU; Kuchner et al. in prep.).  With  F-statistics following \citet{simard11} we compare the residual images of one- and two-component fits to decide which model better represents the light distribution of a particular galaxy. 
To classify the cluster galaxies, we expand the decision tree of \citet{neichel08} \citep[also adopted by others, for example][]{nantais13} to our needs and include the
new capabilities of Galapagos-2 
with varying sets of bulge and disk profiles, as well as color information
and visual classifications. This leads to four distinct classes: compact, peculiar, disk-like, and
objects with a smooth regular light distribution.

Specifically, we follow three main steps for the morphological analysis of the $Zgals$ sample:
i) We perform surface brightness profile measurements with Galapagos-2 using three SUBARU-bands for B/T and n; 
ii) we construct RGB color maps for every GALFIT-M stamp; and
iii) we label each galaxy morphologically based on visual inspection in both the individual images and the color map.
The classification starts with dividing the sample of star-forming galaxies into three broad B/T bins (B/T<0.5, 0.5<B/T<0.8, and B/T>0.8) with $\sim$3\% incorrect measurements in the general comparison to HST in the overlap region. Each stamp is then visually inspected for spiral features, color gradient, and offsetting nuclei \citep[see][for details]{nantais13}. %
Owing to the low resolution of the SUBARU imaging, only the most striking spiral features are seen and a smooth appearance is common. Therefore, we call high B/T objects without any visible features \textit{smooth}. It  should be remembered, however, that these are still star-forming galaxies with a prominent bulge and no obvious signs of spiral structures. Galaxies with effective radii $r_{e}$< 3 kpc (corresponding to less than 0.55\,arcseconds at $z \sim 0.4$) are classified as \textit {compact}. Low B/T (or low S\'ersic index) galaxies and those with spiral features are classified as \textit{disk} galaxies, while galaxies with visible disturbances in their stellar structure are called \textit{peculiar}.


\subsection{SFRs}
\label{sec:SFRs}

~~~One of the most reliable and well calibrated SFR indicators is the
\Ha\, EL.
The conversion of \Ha\, luminosity to SFR requires several steps.
To correct for slit losses for $Zgals$ MAC0416 cluster galaxies 
we convolve each VIMOS spectrum  with the SUBARU Cousins R-band filter and then compare
this magnitude with the SUBARU R-band magnitude of the respective galaxy. 
The difference between the two magnitudes gives the aperture
correction factor for each spectrum. 
This correction 
assumes that the  \Ha\, line flux and the R-band continuum suffer equal slit 
losses and that the EW of \Ha\, is constant throughout the entire galaxy. %
Thus, revised  \Ha\, line luminosities $\rm{L}(\rm{H}\alpha)$ are corrected for extinction based on the Balmer decrement and then transformed into SFRs by applying  the \citet{ken98} conversion: $\rm{SFR} (M_{\odot}\rm{yr}^{-1}) = 7.9 \times 10^{-42}
\rm{L}(\rm{H}\alpha)\rm{ergs/s}$.

The SFRs of the zCOSMOS sample at $z \sim 0.4$ and of SDSS galaxies are also computed from  extinction corrected \Ha\, luminosities. 
For the zCOSMOS galaxies an aperture correction based on the observed I-band magnitude was applied to each VIMOS spectrum  to take into account slit losses, as described in  \citet{maier09}. The SDSS SFRs were corrected for fiber losses using the aperture corrections given by
\citet{brinchmann04}, corrections which work for $z>0.04$, as
demonstrated by \citet{kew05}.


\section{Results: Mass-metallicity and mass-sSFR relations in field and cluster}


~~~We discuss several aspects of galaxy evolution in different environments, each presented in the framework of the relations between  mass and sSFR (Sect.\,\ref{SSFRMsel}) and mass and metallicity (Sect.\,\ref{sect:MZR}). We present evolutionary effects between redshifts $z \sim 0.4$ and $z \sim 0$, establishing the MZR at $z \sim 0.4$. We then subdivide the samples to study effects of the cluster and field environments. Finally, we include a more detailed investigation of morphology dependencies of each relation. For a meaningful investigation and  considering our relatively small sample of targets, we divide the galaxies into three (physically motivated) mass-bins:

i) Galaxies with $\rm{M}_{\rm{high}}$ ($\rm{log(M/M_{\odot})}>10.2$)  lie above a threshold invoked by chemical downsizing in the mass-metallicity plane, manifested in the flattening of the MZR  at high stellar masses \citep[cf.][]{maier15}.
Furthermore, this high-mass regime  is controlled by early-type and bulge-dominated galaxies, which explains the low number statistics for our cluster star-forming galaxies (see also Fig.\,\ref{Selec0416});

ii) The intermediate-mass bin   $\rm{M}_{\rm{med}}$ ($9.2<\rm{log(M/M_{\odot})}<10.2$) is well populated by both field and cluster galaxies; 

iii) Galaxies below $\rm{log(M/M_{\odot})} \sim 9.2$, a stellar mass regime that we call $\rm{M}_{\rm{low}}$, will be considered low-mass objects. We do not have comparison galaxies in the field sample for this mass range owing to zCOSMOS-bright selection criteria.


\subsection{ Mass$-$sSFR relation at $z \sim 0.4$}
\label{SSFRMsel}

~~~ The sSFR has been found to be a tight but weak function of mass
at all epochs up to $z \sim 2$ for most star-forming galaxies
(\emph{main sequence, MS}), and to
evolve 
with time increasing by a factor of 2$-$3 to $z\sim0.4$ compared to $z \sim 0$ 
\citep[e.g.,][]{noeske07,elbaz07}.
\citet{peng10} derived a formula for the evolution of the sSFR as a function of mass and time that we use to calculate the mean sSFR(M) at $z \sim 0.4$. For this, we assume a dependence of sSFR on mass as observed for local SDSS galaxies and revised by \citet{renzpeng15}, sSFR$\propto$M$^{\beta}$ with $\beta=-0.24$ (Fig.\,\ref{SSFRMassz07Env}, oblique solid black line).  
The local MS relation was found  to have a dispersion of a factor of 0.3\,dex 
about the mean relation \citep[e.g.,][]{salim07,elbaz07,peng10}, 
indicated by the gray area 
around the mean MS relation at $z \sim 0.4$ in Fig.\,\ref{SSFRMassz07Env}.


~~~ The distribution in the mass$-$sSFR plane of our sample of field and cluster galaxies at $z\sim 0.4$  is comparable (Fig.\,\ref{SSFRMassz07Env}), in particular in the stellar mass bin $\rm{M}_{\rm{med}}$ 
where the $Zgals$ sample has a uniform coverage with respect to the parent sample  (see Fig.\,\ref{Selec0416}). The median values of the two samples {in this mass range} and the scatter of the data points are also  consistent with the mean MS relation and scatter predicted by \citet{peng10} at $z \sim 0.4$.
In the lower mass bin  $\rm{M}_{\rm{low}}$
there are no field galaxies from zCOSMOS  since  the zCOSMOS-bright sample was selected to have $I<22.5$, so we cannot study environmental effects on SFRs for lower masses.
In contrast, in the highest mass bin  $\rm{M}_{\rm{high}}$
there are only six cluster galaxies in our sample,  most of them showing lower sSFRs than typical MS galaxies at these redshifts.


Investigating in detail the morphological distribution in our sample of cluster galaxies, a segregation is revealed between bulge-dominated on the one hand and disk and peculiar galaxies on the other (Fig. \ref{SSFRMassOffsets}). We measure a difference between the populations by comparing deviations of the median residuals of the fits.
We find that cluster galaxies with prominent bulges lie systematically lower (>$1 \sigma$) than the disk star-forming galaxies that are closely compatible with the MS. 
Stars in galactic bulges at $z \sim 0.4$ are dominated by an old, red, and passive population \citep{pere13}, accounting little for the overall ongoing star formation of the galaxy. As a consequence, our galaxies in the smooth category,  which have rather high B/T values, are expected to display lower sSFRs.
Peculiar galaxies on the other hand lie above the MS indicating an increase in current star formation with notable scatter. This type of galaxy is nearly impossible to distinguish through classifications based on single S\'ersic $n$ and were therefore not  captured in some previous studies.

Numerous observations describe an increase of bulge-dominated galaxies in high-density regions, such as clusters of galaxies \citep[e.g.,][]{dressler80, kormendy09}. 
The observed slope and scatter of the  mass$-$sSFR relation reflect the increase in bulge-mass fractions in higher mass galaxies and therefore a diversity in SFHs \citep{abram14}.
Our finding is in compliance with \citet{whit15}, who link the scatter and slope of the MS to galaxy structures using single S\'ersic measurements as a proxy for a dominant bulge or disk.  For our work, we perform galaxy decompositions and therefore directly measure the presence and strength of 
bulges.

We see indications of the same trend of lower sSFRs in $\rm{log(M/M_{\odot})}<10.2$ compact galaxies. However, decompositions into bulges and disks become less reliable for these small galaxies in ground-based imaging data (as revealed by our SUBARU - HST comparisons).
Therefore, we can only speculate about the presence of a bulge in compact galaxies. In the local universe, however, Ellison et al. (2008) have  demonstrated a potential link between galaxy sizes and bulge fractions. They found that galaxies with smaller half light radii 
($r_{e} < 3$\,kpc)
have higher B/T ratios (B/T>0.4), which supports our hypothesis. Figure\,\ref{SSFRMassOffsets}  shows similar effects of lower sSFRs for smooth galaxies (bulge-dominated) and to some extent also for compact galaxies on the one hand, and of higher sSFRs for disk and peculiar types (low bulge fractions) on the other hand.
%

\subsection{ MZR at $z\sim 0.4$ in field and cluster}
\label{sect:MZR}

 ~~~Figure\,\ref{MZRz07UK} shows the MZR of  our $Zgals$ sample of 76  MAC0416 $z \sim 0.4$ cluster members (blue symbols)
and 93 zCOSMOS field galaxies (green symbols) compared to SDSS (gray area).
All metallicities were computed with the same method  described in Sect.\,\ref{Oxabund}.
For the matched SDSS sample constructed as described in Sect.\,\ref{sec:SDSS}, we indicate 16th and 84th percentiles, and the medians (50th percentiles) of the distribution of O/H values in the respective mass bin as solid lines
and gray area. We note that the mean SDSS O/H values in the respective mass bins differ from the median O/Hs by less than 0.01\,dex. We also extrapolate the MZR of the matched SDSS sample to lower stellar masses by assuming that the slope of the low-mass MZR stays constant for  $\rm{log(M/M_{\odot})} < 9.2$.
While both the local and distant sample of field  galaxies  have stellar masses $\rm{logM/M_{\odot}} > 9.2$, several $Zgals$ cluster galaxies extend  to lower masses $\rm{log(M/M_{\odot})} \sim 8.5$ thanks to the deep spectroscopy of CLASH-VLT. 

The polynomial fits (third degree) to the data points (blue and green thick lines in the left panel of Fig.\,\ref{MZRz07UK}) and also the median values (right panel) are very similar for cluster and field galaxies of intermediate masses.
We determine an evolution of $\sim 0.1$ dex on average for these $\rm{M}_{\rm{med}}$ galaxies.
This offset for field galaxies towards lower metallicities  
is in agreement with the value given by \citet{perez13}. 
From their Table\,3 of their zCOSMOS sample at similar redshifts, we infer an evolution of $ \sim 0.09$\,dex.


At the highest masses $\rm{M}_{\rm{high}}$ 
most of the zCOSMOS field galaxies  have lower O/Hs than the local ones,
while the high-mass cluster galaxies (albeit with only a few measurements) exhibit higher O/H values, comparable to SDSS metallicities.
This may indicate that the ``chemical downsizing'' effect described in \citet{maier15} regarding the time evolution of the mass threshold below which one finds low metallicities  may have a different mass/time dependence in field and cluster environment. This indicative interpretation has to be verified with a larger sample of metallicity measurements for cluster star-forming $\rm{M}_{\rm{high}}$ galaxies.


In their environmental study of central and satellite SDSS galaxies, \citet{pasq12} report a small metallicity increase of 0.06\,dex at  $\rm{logM/M_{\odot}} \sim 8.5$, which decreases to  0.004\,dex at $\rm{logM/M_{\odot}} > 9.3$, for satellite galaxies in comparison to central galaxies of the same stellar mass. At higher redshifts we do not have the benefit of such a large database of metallicity measurements and restrictions in mass ranges or smaller samples like ours cannot reproduce this accuracy. 

In their publication, \citet{pasq12} speculate whether the 
slight increase of the MZR at lower masses found in satellite galaxies can be explained through strangulation scenarios.
Strangulation prevents satellite galaxies from accreting new, low-metallicity gas which would otherwise dilute their interstellar medium. 
We will  explore 
this scenario further in Sect.\,\ref{sec:FMR} when we discuss the Z(M,SFR) of cluster galaxies.

In a numerical work of environmental effects on metallicities,
\cite{dave11} used cosmological hydrodynamic simulations and compared the MZR of model galaxies subdivided by local galaxy density.
They found that model galaxies in high-density regions lie above the mean MZR by $\sim 0.05$\,dex; however,  the differences disappear at the massive end in all models.
The general trend of both studies, numerical works and local observations, agree with our findings: the MZRs of cluster and field galaxies  differ only marginally.


Several studies in the local Universe found a correlation between gas-phase metallicites and galaxy morphology, such that more bulge-dominated galaxies are more metal rich \citep[e.g.,][]{vila92,zarit93,zari94}. 
Similar trends that connect samples of high concentration galaxies (bulge-dominated)  to higher fractions of solar metallicities in comparison to low-concentration samples (more disk-like)  and lower metallicities were reported by \citet{pasq12}.  
Morphologies in our sample are uniformly distributed over stellar masses and colors within the blue cloud, especially in our intermediate-mass bin $\rm{M}_{\rm{med}}$, and any trends found are therefore unlikely to  result purely from the strong correlations between luminosity, color, and morphology. The chemical evolution of galaxies is connected to the SFH, and is also  influenced by the environment through gas interchange with the surroundings. We therefore expect trends to be intimately linked to the observed scatter of the MS and differences in sSFRs.
Figure\,\ref{fig:MZmorph} shows the median offsets to the local SDSS MZR  for $Zgals$ MAC0416 galaxies, divided by their morphologies.  Just like in the mass$-$sSFR relation, we find different behaviors for smooth/compact vs. disk/peculiar types. 
In agreement with predictions from the FMR, the smooth and compact galaxies that have lower sSFRs display higher metallicities 
with mean values only slightly below, and in some cases comparable to (within 1$\sigma$), the mean local MZR. In contrast, disk-dominated and peculiar objects (with higher sSFR) exhibit metallicities well below the local MZR. 

We note that,  as discussed in Sect.\,\ref{SSFRMsel}, smooth and compact galaxies have higher bulge fractions than disk-dominated and peculiar objects. Simulations have shown that bulges appear to stabilize against inflows of metal-poor gas and nuclear starbursts driving galactic outflows \citep[e.g.,][]{miho94, miho96, cox08}. Bulge-dominated galaxies like the \textit{smooth} galaxies in our $Zgals$ sample therefore possibly retain their metal-enriched gas more easily. For the other types, like \textit{disk} and \textit{peculiar} $Zgals$ galaxies, inflowing gas may be funneled more efficiently to their centers with low bulge fractions.
%

\section{Discussion: Environmental effects and evidence for starvation}
\label{sec:discussion}

~~~ The location of galaxies in projected phase-space (cluster-centric radius vs. line-of-sight velocity relative to the cluster redshift) provides valuable information on their accretion history: on average, infalling galaxies are spatially separated from virialized galaxies. 
Simulations of mass assembly around clusters have helped to identify these regions in phase-space.
Stacked caustic (isodensity profile) plots of galaxies orbiting the 30 most massive clusters in the Millenium simulation demonstrate how galaxies of different accretion histories populate different areas in a characteristic trumpet shaped caustic profile \citep{haines15}. Recently accreted galaxies (with, in general, higher line-of-sight velocities) are located outside these caustics and are thus well separated from the dense cluster core where galaxies 
are more likely to have been accreted at  earlier times.

For our analysis, we 
define
the rest-frame velocity as $v_{rf} = c (z-z_{mean})/(1+z_{mean})$, where $z_{mean}=0.3972$ is the mean cluster redshift, and we depict R vs. $|v_{rf}|$ in Fig.\,\ref{fig:PhaseSpace}, with R being the projected radial distance from the cluster center. To separate mass and environmental effects, we concentrate on a limited range in stellar mass and consider only 53 MAC0416 cluster galaxies with $\rm{M}_{\rm{med}}$ ($9.2<\rm{log(M/M_{\odot})}<10.2$).
The dashed magenta line in Fig.\,\ref{fig:PhaseSpace}  indicates the caustic determined by B16,  separating the regions in phase-space of recently accreted and infalling
galaxies (outside caustics) from the region of the galaxies accreted longer  ago and a possibly virialized population \citep[using the nomenclature of
][see also \citet{jaffe15}]{haines15}. For simplicity, we  call the two classes ``infalling'' (above and to the right of the magenta dashed line, star symbols) and ``accreted'' (filled circles).

We note that only 5 out of 18 infalling  $\rm{M}_{\rm{med}}$ galaxies (28\%) have higher O/Hs consistent with the local MZR, while this fraction is higher for accreted  $\rm{M}_{\rm{med}}$ objects, namely 22 out of 35 galaxies (63\%).
As discussed and shown in the following, infalling galaxies with lower metallicities than the local MZR (blue star symbols, representing 72\% of the infalling population) can increase their O/Hs by $\sim 0.2$\,dex in   $t_{\rm{enrich}}\sim 1$\,Gyr reaching the local MZR while they move towards the central region of the cluster.
This enrichment time 
is shorter than the crossing time $t_{\rm{crossing}}=R_{200}/\sigma$ of the cluster, which is about 2\,Gyrs,  and also smaller than the $2-3$\,Gyrs needed by galaxies in the infalling region to reach the central, denser part of the  cluster, as derived by \citet{haines15}. 
This makes a strangulation scenario a plausible mechanism, in which the gas metallicity increases because the reservoir of pristine gas is depleted when the galaxies enter the cluster. We explore additional evidence for this scenario in the following sections. 
%


\subsection{The Z(M,SFR) of  cluster galaxies at $z \sim 0.4$}
\label{sec:FMR}

~~~In order to explain star formation as a second parameter in the MZR, some authors have ascribed an ad hoc inflow of gas as the  driver responsible for the  dilution of metallicity and increase in SFR \citep{mannu10,dave12,dayal13}.
A different explanation was put forward by Li13, who proposed a simple model of 
galaxy evolution in which the SFR is regulated by the mass of gas present in a galaxy, implying that Z depends on both stellar mass and current SFR.
They derived a Z(M,SFR) relation 
that depends on the internal parameters describing the regulator system, specifically the star formation efficiency $\epsilon = \rm{SFR}/\rm{M}_{gas}$
 and the mass-loading of any wind that drives gas out of the system, their $\lambda = outflow / \rm{SFR}$.  
The local SDSS data from \citet{mannu10}
can be reproduced with astrophysically plausible values of $\epsilon(M)$ and $\lambda(M)$
(see  Eq.\,40 in Li13 and   Eq.\,3 in \citet{maier14}), Z$_{eq}=Z_{0}+y/(1+\lambda(1-R)^{-1}+\epsilon^{-1}((1+\beta-b)SFR/M-0.15))$,
where $Z_{eq}$ is the equilibrium value of the metallicity, $Z_{0}$ is the metallicity of the infalling gas, $y$ the yield, and $R$ the returned fraction.
The possible values of the different parameters $\epsilon$, $\lambda$, $y$, and $b$ are given in Table\,1 of Li13.
These analytic approximations 
have since been verified by more detailed calculations in \citet{pipino14}.

Our aim is to challenge this model description of a fundamental relation in a dense cluster environment at higher redshift,
i.e., we want to investigate whether the Z(M,SFR) dependence is similar to the one found in the local  Universe so that the FMR is still valid.
To this purpose we calculate the expected O/H values from the Li13 model for each galaxy individually with their respective stellar mass M and SFR. 
We then compare the predicted values with the measured ones and show the difference  in Fig.\,\ref{FMRDiff}.
We find that 87\% of the zCOSMOS field comparison galaxies have O/H values that agree with the model expectations within 0.2\,dex, the nominal maximum error in Z determination. 
We conclude therefore that the majority of zCOSMOS field galaxies at $z \sim 0.4$ (with $\rm{log(M/M_{\odot})}>9.2$) obey the FMR. This is in accordance with 
\citet{cresci12} who used a small subsample 
of zCOSMOS $z \sim 0.4 $ galaxies, albeit  with a strong S/N selection: $S/N>20$ and $S/N>8$ in \Hb\, flux for log(M/\msun )$>10$ and log(M/\msun )$<10$, respectively. They also considered the difference between measured O/H and predicted O/H measurements, but used the \citet{mannu10} predictions instead of the Li13 ones to conclude that the FMR holds for field galaxies at $z \sim 0.4$. 


To investigate the FMR for cluster members, we show in panel c) of
Fig.\,\ref{FMRDiff} the difference in O/H measurements from predictions using both primordial and enriched inflowing gas. Eighty-eight percent of $Zgals$ cluster galaxies with 
$\rm{log(M/M_{\odot})}>9.2$ have O/H values that are in accordance with the model expectations within 0.2\,dex, similar to the behavior of zCOSMOS field galaxies.
In contrast, lower mass $\rm{M}_{\rm{low}}$ cluster members deviate systematically from the FMR 
predictions.
For the first time we can explore the Z(M,SFR) of star-forming cluster galaxies at higher redshifts down to low masses $\rm{M}_{\rm{low}}$ ($8.5<\rm{logM/M_{\odot}}< 9.2$) with CLASH-VLT.
These $\rm{M}_{\rm{low}}$ cluster galaxies are not compatible with the FMR if  primordial inflowing gas ($Z_{0}=0$) is assumed (open circles), having metallicities two to three times higher than predicted.  We note that most of these low-mass galaxies are accreted objects (except three infalling $\rm{M}_{\rm{low}}$ objects shown in blue). This suggests that galaxies of lower masses $\rm{M}_{\rm{low}}$ are more prone to environmental effects when they are accreted by the cluster owing to their lower gravitational potential.
Better agreement with the FMR can be achieved with infall of enriched gas with metallicity  $Z_{0}/y=0.1$ (filled symbols).
This is in compliance with the predictions of \citet{pengmaio14} that the inflowing gas should become progressively metal enriched ($Z_{0}$ is higher) in dense regions, producing higher metallicities for cluster galaxies of low masses compared to similar field galaxies. \citet{pengmaio14} do not give a physical explanation how metal-enriched gas inflow can  happen in the dense cluster environment.
However, they offer an alternative scenario based on strangulation, in which  the inflow of pristine gas is stopped, as a possible explanation for higher O/Hs in lower mass objects. They do not elaborate this scenario further because they find, as we do,   that the sSFRs are similar in different environments  (see Fig.\,\ref{SSFRMassz07Env}).

It should be noted that we call strangulation/starvation a scenario in which the supply of cold gas onto the galaxy disk is halted because the halo gas is stripped owing to external forces. In this case star formation can continue, using the gas available in the disk until it is completely used up.
Therefore, similar sSFRs for star-forming cluster and field galaxies do not completely exclude strangulation as a plausible mechanism for the chemical enrichment effects we see in the cluster galaxies.
In a context of a rich cluster, this scenario was first described by \citet{larson80}, who used the removal of the gas reservoir due to external forces to explain the passive cluster galaxy population.

To explore the strangulation scenario, we divide the $\rm{M}_{\rm{med}}$  cluster sample into infalling and accreted objects according to their position in the phase-space diagram (Fig.\,\ref{fig:PhaseSpace}) and depict them in blue and red
in Fig.\,\ref{FMRDiff}.  The difference between measured and predicted O/Hs for most of these $\rm{M}_{\rm{med}}$ cluster objects is very similar  for the two scenarios with inflow of primordial gas (open circles) and enriched gas (filled circles): the difference between the values of the open and the respective filled circles for most $\rm{M}_{\rm{med}}$ galaxies is less than $\sim 0.02$\,dex. Therefore, the histograms of $\rm{M}_{\rm{med}}$ galaxies shown in panel a) of Fig.\,\ref{FMRDiff} are very similar for the two scenarios of inflowing enriched or primordial gas.
These histograms show that the  O/H distribution of  $\rm{M}_{\rm{med}}$ infalling
cluster galaxies 
is in  good agreement with the FMR, while accreted galaxies have a distribution shifted to higher  metallicities, in general higher than predicted by the Li13 model.
This indicates that galaxies may increase their metallicities as a result of starvation after being accreted by the cluster. On the other hand, infalling galaxies are still star-forming, with their  O/Hs  not yet influenced by environmental effects of the cluster, in general following the FMR and behaving like field galaxies (panel b of Fig.\,\ref{FMRDiff}).  
In a recent paper, \citet{peng15} used the SDSS local sample and found that the \emph{stellar} metallicity of satellite galaxies is slightly higher than that of central galaxies for stellar masses $\rm{log(M/M_{\odot})}<10.2$ (value converted to a Salpeter IMF as used in this paper). They inferred that lower mass satellite galaxies are more prone to  strangulation, implying that  environmental effects are responsible for stopping the inflow of gas. We now find additional signs of strangulation due to environment at $z \sim 0.4$ for galaxies with similar masses,  $\rm{log(M/M_{\odot})}<10.2$.
In the next section, we discuss further evidence that strangulation is a plausible mechanism for explaining the metallicities of cluster galaxies.


\subsection{Co-evolution of SFR and Z}
\label{sect:MZRpaths}

~~~To understand the combined  SFR and metallicity evolution of galaxies, we consider four toy models of different SFHs:
a constant SFR and primordial gas inflow; an exponentially declining SFR with a timescale $\tau$ of 3\,Gyrs and inflow; and  strangulation scenarios, i.e., without any inflow, for the two SFHs. 
We follow the evolution  of the model galaxy in these four scenarios for  3\,Gyrs, which roughly corresponds to i) the time between the redshift of the MAC0416 cluster, $z \sim 0.4$, and the upper redshift range of the SDSS comparison sample, $z \sim 0.08$, and ii) the time  span between $z \sim 0.8$ and $z \sim 0.4$. The four models allow us to follow i) the evolution of an infalling low metallicity galaxy (blue star symbols in Fig.\,\ref{fig:PhaseSpace}) that will reach more inner regions of the cluster and increase its metallicity, as expected from the observed O/Hs at $z \sim 0.4$ of most accreted cluster members,
and ii) the evolution of a galaxy infalling into the cluster at $z \sim 0.8$ and becoming an accreted cluster galaxy when being observed at $z \sim 0.4$, likely becoming a higher metallicity object (red filled circles in Fig.\,\ref{fig:PhaseSpace}). Thus, the paths of our model galaxies are assumed to start in a region populated by infalling low-metallicity galaxies, and are expected to end in the region of the mass-metallicity diagram populated by higher metallicity objects (with O/Hs consistent with the local MZR, as indicated by the red dash-dotted line and red symbols in panel b of Fig.\,\ref{fig:DeltaOHDeltaSSFR}).

As discussed in the previous section, Fig.\,\ref{FMRDiff} demonstrates that the more massive ($\rm{log(M/M_{\odot})}>9.2$) field and \emph{infalling} cluster galaxies at $z \sim 0.4$ are generally compatible with the Z(M,SFR) relation as given by Li13 (their Eq.\,40). 
Since this equation, which we reproduced in Sect.\,\ref{sec:FMR}, is by construction also applicable to local SDSS galaxies, we can assign a SFR value to every grid point in the MZR diagram.
In the following we use Z(logM,SFR) to denote the metallicity O/H as a function of stellar mass and SFR.

The starting point of the evolutionary tracks for the four model galaxies at $z=0.4$ (shown as a filled green circle with a black cross in  panels b and c of Fig.\,\ref{fig:DeltaOHDeltaSSFR}) was derived by extrapolating the local MS to this redshift, following the evolution of the sSFR, derived in \citet{peng10}. Initial values for stellar mass and SFR are chosen as 
$\rm{log(\rm{M}_{\rm{begin}}/M_{\odot})}=9.5$ and $\rm{SFR}_{\rm{begin}}=1.2 M_{\odot}/yr$. 
Following considerations of the FMR (Li13), the model galaxy starts with an initial metallicity value of 
$\rm{Z}_{\rm{begin}}(9.5,1.2)=8.55$. 
These  are typical (mean) $\rm{Z}(log\rm{M},\rm{SFR})$ values  for infalling, low metallicity, MS $Zgals$ cluster galaxies with $9.2<\rm{log(M/M_{\odot})}<9.8$ (blue star symbols in Fig.\,\ref{fig:DeltaOHDeltaSSFR}).  
On the other hand, the five cluster  infalling galaxies with higher metallicities show, with one exception, lower sSFRs than the MS (red star symbols in panel c of Fig.\,\ref{fig:DeltaOHDeltaSSFR}). The lower sSFRs (and higher O/Hs) of these \emph{infalling} galaxies could be due to pre-processing  in the group environment \citep[e.g.,][]{zabmul} prior to arriving in the infall region of the cluster.

We follow the tracks of our model galaxies in the MZR and in the $\Delta \rm{sSFR} -\Delta \rm{O/H}$ diagrams.
 We chose $\Delta \rm{sSFR}$ to be the offset to the MS at $z \sim 0.4$ (see Fig.\,\ref{SSFRMassOffsets}), while  $\Delta \rm{O/H}$ is the offset between the measured O/H and the local SDSS MZR value for O/H at the respective mass (see Fig.\,\ref{fig:MZmorph}).
The sSFR offset from the MS at a given mass indicates the level of star formation compared to what is typical for a galaxy of that mass.


The magenta lines in panel b of Fig.\,\ref{fig:DeltaOHDeltaSSFR} show the Li13 expectations of the Z(M,SFR) for galaxies on the MS at different redshifts, with dashed, solid, and dotted lines corresponding to $z \sim 0.8$, $z \sim 0.4$ and $z \sim 0.06$ MS galaxies.
A MS model galaxy with $\rm{Z}_{\rm{start}}(9.5,1.2)=8.55$ lies on the MZR of MS galaxies at $z \sim 0.4$ (solid magenta line).  It forms stars for 3\,Gyrs with a constant SFR and increases its stellar mass to $\rm{log(\rm{M}_{\rm{end}}/M_{\odot})}=9.83$, and its metallicity to $\rm{Z}_{\rm{end}}(9.83,1.2)=8.78$ (cyan track). 
Although the SFR is constant, the sSFR of the model galaxy with constant SFR  decreases 
because stellar mass is formed, reaching the  expected MZR of MS galaxies  at $z\sim 0.06$ (dotted magenta line in the panels a and b of Fig.\,\ref{fig:DeltaOHDeltaSSFR}).
Thus, we note that a MS $z \sim 0.4$ model galaxy with constant SFR can reach the MS of SDSS galaxies without decreasing its SFR, but simply because it 
increases mass through constant SFR, which leads to a decrease in its sSFR.

The tracks for a model galaxy forming stars with an exponentially declining SFR on a timescale $\tau$ of 3\,Gyrs and having an inflow of gas are shown by the green tracks in panels a and b of Fig.\,\ref{fig:DeltaOHDeltaSSFR}. In this case $\rm{log(\rm{M}_{\rm{end}}/M_{\odot})}=9.73$, $\rm{SFR}_{\rm{end}}=0.44\,\rm{M}_{\odot}$/yr, and the sSFR decreases  after 3\,Gyrs by a larger amount than in the case of constant SFR
(panel a), and the model galaxy reaches $\rm{Z}_{\rm{end}}(9.73,0.44)=8.78$, the same metallicity as in the case of constant SFR (panel b). 
Any naively assumed difference in $\rm{Z}_{\rm{end}}$ (galaxies with lower $\rm{M}_{\rm{end}}$ should also have lower $\rm{Z}_{\rm{end}}$) cancels out because smaller $\rm{SFR}_{\rm{end}}$ implies higher $\rm{Z}_{\rm{end}}$.
%


In our assumption of beginning strangulation, stars continue to form and consume the available gas in the disk  either with  a constant SFR (blue tracks in Fig.\,\ref{fig:DeltaOHDeltaSSFR}) or with a slowly decreasing SFR on a timescale $\tau$ of 3\,Gyrs (black tracks in Fig.\,\ref{fig:DeltaOHDeltaSSFR}).
We find that a model galaxy with an exponentially faster declining SFR with a $\tau=1$\,Gyr becomes passive by 
 $z \sim 0.06$ with a sSFR about 30 times lower than at $z \sim 0.4$, therefore disappearing from panel a of Fig.\,\ref{fig:DeltaOHDeltaSSFR}. 
Recent results of \citet{cantale16} strengthen the plausibility of our toy models with ongoing star formation  for several Gyrs since they found that galaxies are able to continue forming stars for up to 5\,Gyrs after being accreted into clusters. Additionally, \citet{jaffe15} studied HI gas masses in cluster galaxies at $z \sim 0.2$ and found a  plausible scenario in which galaxies affected by ram-pressure stripping as they first fall into the cluster are able to continue to form stars in the remaining gas disk.

In the strangulation scenario, our model galaxy reaches the same values of $\rm{log}\rm{M}_{\rm{end}}$ and $\rm{SFR}_{\rm{end}}$ as above, but the increase in metallicity is higher than in the non-strangulation case, i.e., $\rm{Z}_{\rm{end}}$ is higher.
After stopping the inflow of gas, the evolution of the galaxy can be approximated with a closed-box model. This first-order assumption neglects effects of gas outflows that continue after the supply of cold gas from the surroundings comes to a halt.
However, studies  have shown \citep[e.g.,][]{peng15} that in the case of truncated gas inflows, metallicities are independent of outflows, because the element mix of the retained gas stays constant. Thus, a closed-box model is still a good approximation.
\citet{maier06} constructed a large grid of P\'egase2 models to explore which region of the
parameter space could reproduce the constraints imposed by the local
metallicity--luminosity relation and by the metallicities and luminosities of galaxies at
higher redshifts. 
They explored
models in which the gas inflow on the model galaxy stops after about 1\,Gyr.
 The galaxy subsequently forms stars in a 
``closed-box''-like scenario.
The tracks
of these closed-box models show that, after gas inflow stops, the model galaxies increase their metallicities by $\sim 0.2$\,dex in $t_{\rm{enrich}} \sim 1$\,Gyr and  by more than a factor of two, namely by $\sim 0.35$\,dex, in $t_{\rm{enrich}} \sim 3$\,Gyr.


In both strangulation scenarios, we therefore assume a metallicity increase of more than a factor of two in $t_{\rm{enrich}}\sim 3$\,Gyr, as justified by the P\'egase2 closed-box models, which leads to final metallicity values of
 $\rm{Z}_{\rm{end}}=8.9$ (black and blue tracks in Fig.\,\ref{fig:DeltaOHDeltaSSFR}).  
These  $\rm{Z}_{\rm{end}}$ values of the model starvation paths lie in the region of the mass--metallicity diagram populated by higher metallicity SDSS objects (the metallicity of the red symbols in panel b in Fig.\,\ref{fig:DeltaOHDeltaSSFR} are in agreement with the local SDSS MZR). This indicates that starvation is a plausible scenario for the evolution of the metallicities of $z \sim 0.4$ infalling galaxies to higher metallicities  after 3\,Gyrs, consistent with the upper, higher metallicity part of the $z \sim 0.06$ MZR. 
Additionally, the $\rm{Z}_{\rm{end}}$ values  of the model starvation paths are  also in agreement with the O/Hs of several accreted cluster galaxies at $z \sim 0.4$ (red filled circles in Fig.\,\ref{fig:DeltaOHDeltaSSFR}).
Therefore, $z \sim 0.4$  accreted cluster members with higher metallicities  are likely to have been infalling into the cluster at earlier times ($z \sim 0.8$)   and to have suffered strangulation in the time span of about 3\,Gyrs between $z \sim 0.8$ and $z \sim 0.4$.


\section{Summary}

\begin{figure*}
\includegraphics[width=8cm,angle=270,clip=true]{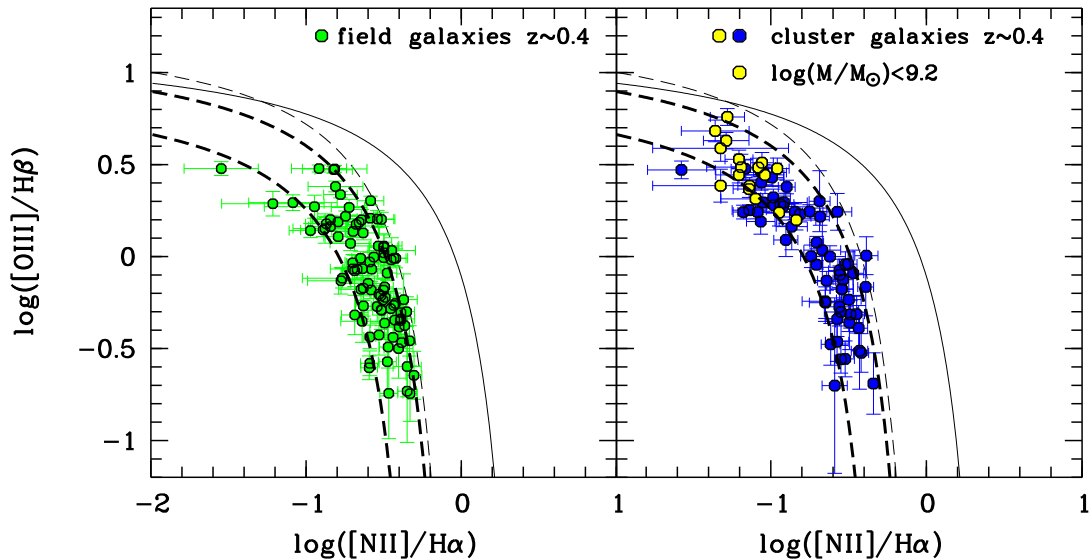}
\caption
{
\label{fig:BPT} 
\footnotesize 
BPT \citep{bald81} diagnostic diagram used to distinguish  star formation-dominated galaxies
from AGNs. The left diagram shows the field galaxies at $z\sim 0.4$, and the right panel  the $Zgals$ cluster members of MAC0416  (cluster galaxies of similar high masses to the field comparison sample are depicted in blue).
Both the field and cluster galaxies at $z \sim 0.4$
lie under and to the left of the theoretical curve (solid)  of \citet{kewley01} and of the empirical curve  (dashed)  of \citet{kaufm03},
 which separate star-forming galaxies (below and left of the
curves) from AGNs (above and right of the curves),  indicating that in all of
them the dominant source of ionization in the gas is recent star formation.
Most galaxies lie on the star-forming metallicity sequence with rough boundaries indicated by the thick dashed lines.
}
\end{figure*}

~~~This study of environmental effects on the Z(M,SFR) relation is based on VIMOS spectroscopy of cluster galaxies in the MAC0416 cluster, supplemented by a zCOSMOS field galaxy sample at similar redshifts $z \sim 0.4$.%
The main results can be summarized as follows:

1. The star-forming cluster galaxies at $z \sim 0.4$ are not dominated by AGNs, and they follow the star-forming abundance sequence in the BPT diagram (Fig.\,\ref{fig:BPT}).

\begin{figure*}
\includegraphics[width=17cm,angle=0,clip=true]{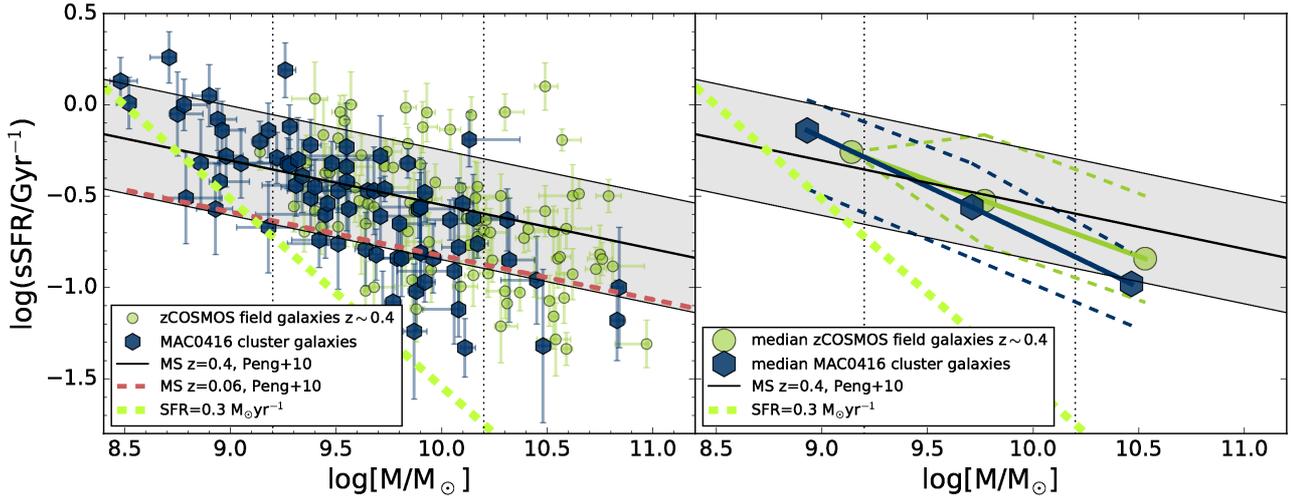}
\caption
{
\label{SSFRMassz07Env} 
\footnotesize 
Left panel: The mass$-$sSFR relation for MAC0416 cluster galaxies (blue symbols) and zCOSMOS field galaxies (green) at $z \sim 0.4$. 
The oblique solid thick black line shows the MS at $z \sim 0.4$ and its dispersion (indicated by the gray area), and the dashed red line depicts the local SDSS  MS, using Eq. 1 in \citet{peng10} for $z \sim 0.4$ and  $z \sim 0.06$, respectively. 
Right panel: the median values of the mass$-$sSFR relations at $z \sim 0.4$ are shown as blue (cluster) and green (field) solid lines and big symbols, while the corresponding dashed oblique lines enclose 68\% of the data. 
The dashed thick light green line shows the completeness limit of our $Zgals$ cluster sample, $\rm{SFR} \sim 0.3\,\rm{M}_{\odot}$/yr.
}
\end{figure*}

\begin{figure*}
\includegraphics[width=18cm,angle=0,clip=true]{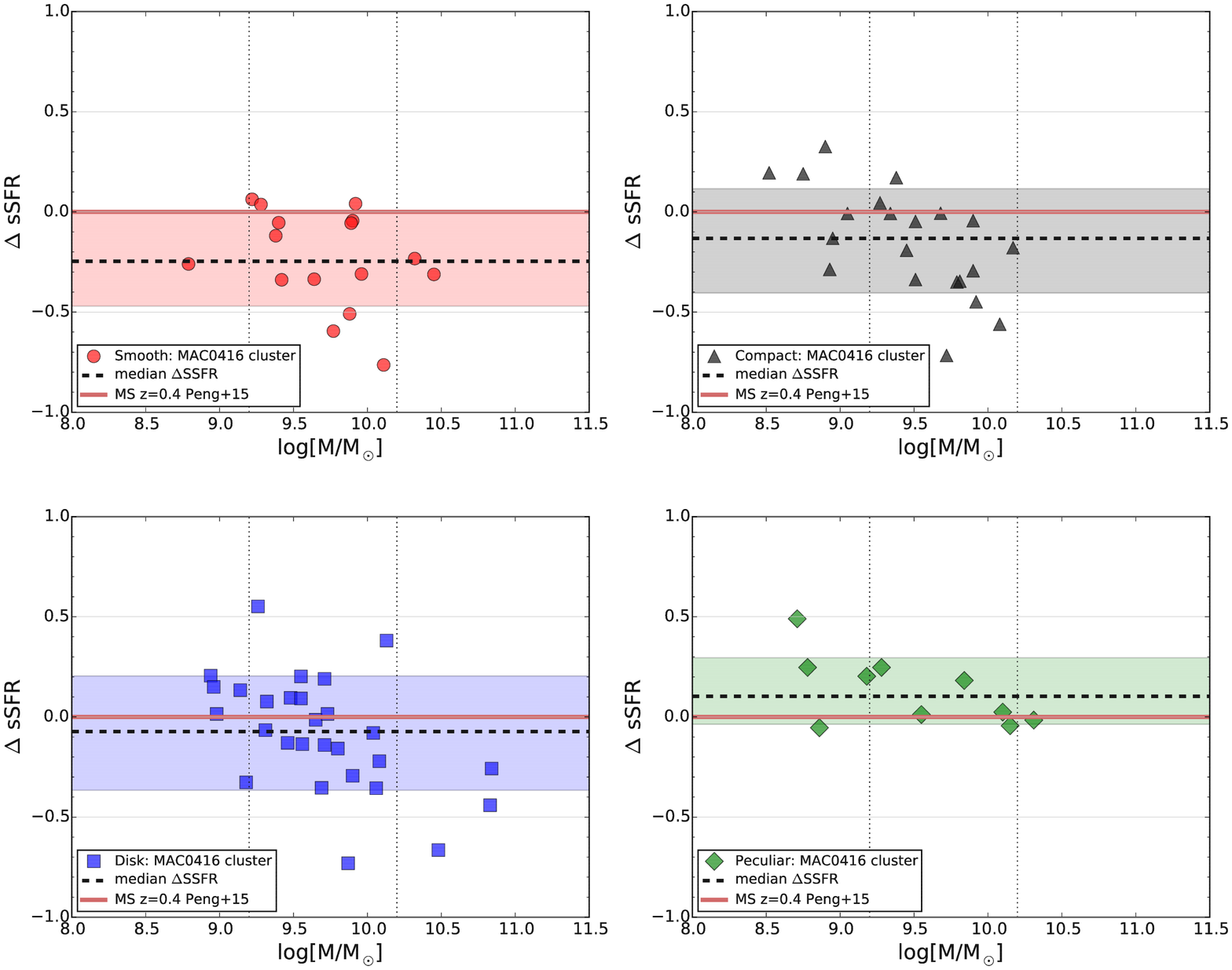}
\caption
{
\label{SSFRMassOffsets} 
\footnotesize 
 Offsets of cluster galaxies with different morphologies from the MS at $z \sim 0.4$. Blue squares are disks, red circles smooth, black compact, and green irregulars. The red solid lines show the MS at $z \sim0.4$ predicted by \citet{peng10}. The shaded gray area is the $1 \sigma$ scatter of the respective morphological category, while the dashed black solid line is the median value. 
}
\end{figure*}

\begin{figure*}
\includegraphics[width=17cm,angle=0,clip=true]{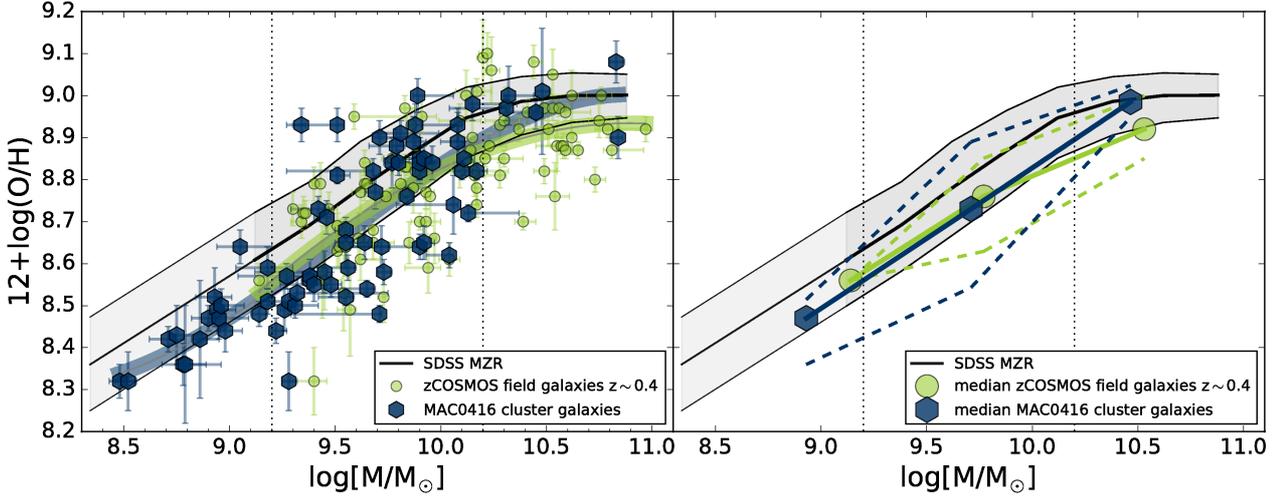}
\caption
{
\label{MZRz07UK} 
\footnotesize 
Left panel: MZR for 76 MAC0416 cluster members ($Zgals$ sample, blue symbols) and 93 field galaxies (green symbols) at $z\sim 0.4$. Polynomial fits to the data points (blue and green thick lines) show that the MZRs for cluster and field galaxies are similar at intermediate masses  $\rm{M}_{\rm{med}}$,
 where the $Zgals$ sample has a uniform coverage with respect to the parent sample  (see Fig.\,\ref{Selec0416}).
Right panel:  Median values of MZR at $z \sim 0.4$ are shown in blue (cluster) and green (field), with the corresponding dashed oblique lines enclosing 68\% of the data. 
In both panels we extrapolated the MZR of the SDSS sample to lower stellar masses by assuming that the slope of the low-mass MZR stays constant for  $\rm{log(M/M_{\odot})} < 9.2$.
}
\end{figure*}

\begin{figure*}
\includegraphics[width=17cm,clip=true]{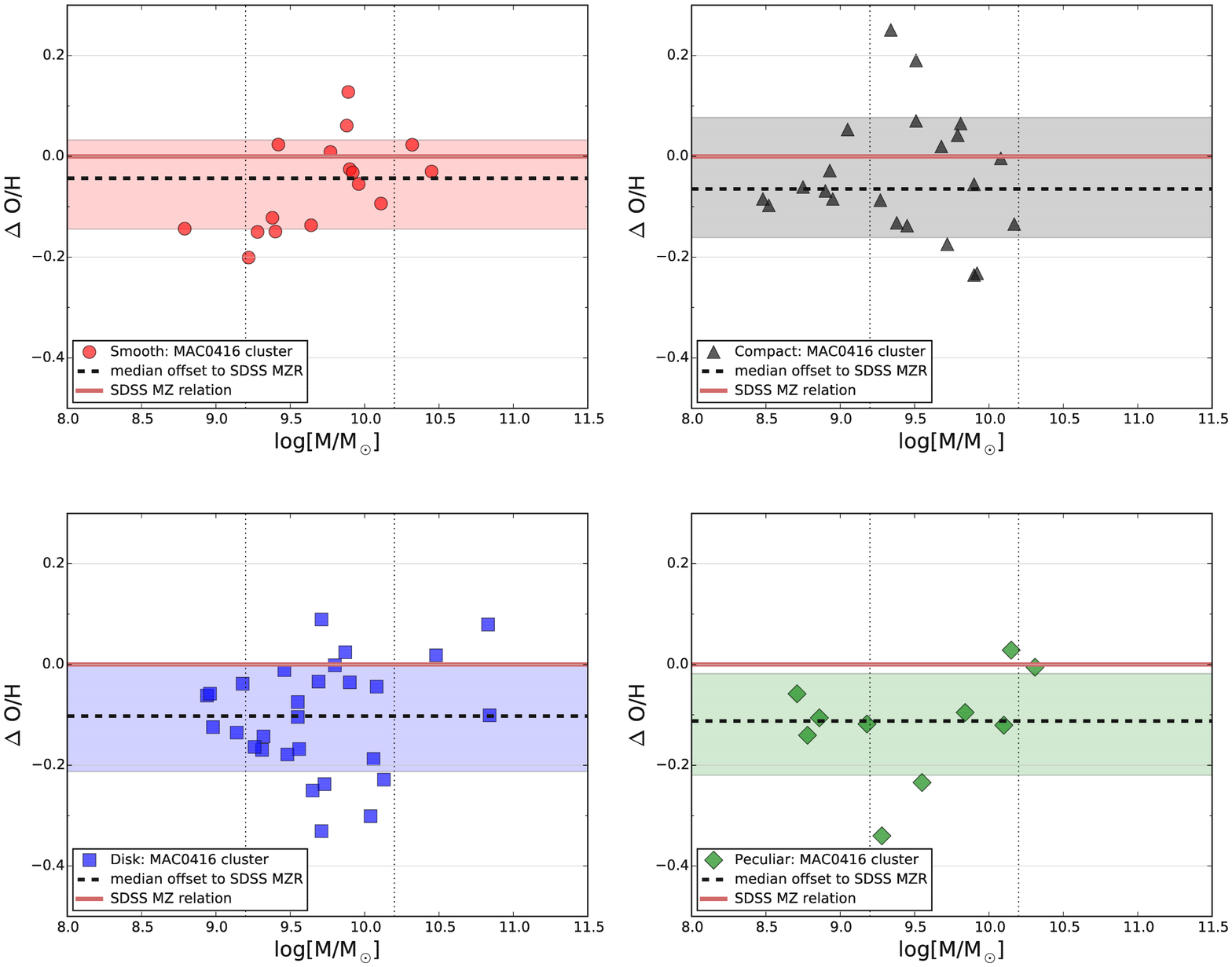}
\caption
{
\label{fig:MZmorph} 
\footnotesize 
Median offsets to the local relation for $Zgals$ MAC0416 galaxies, divided by their morphologies. We find different behaviors for smooth/compact vs. disk/peculiar types with respect to the local relation. While for the latter we measure O/H abundances that are lower than the local relation, those galaxies with more prominent bulges and compact galaxies are more enhanced in gas metallicities. 
}
\end{figure*}

2. Cluster and field galaxies at $z \sim 0.4$ show, on average, similar MZR and mass$-$sSFR relations for stellar masses $9.2<\rm{log(M/M_{\odot})}<10.2$ (Figs.\,\ref{SSFRMassz07Env} and \ref{MZRz07UK}), in accordance with the findings of numerical studies and local observations.
On the other hand, we find that bulge-dominated (smooth) and compact cluster galaxies  have lower sSFRs and higher O/Hs than  their disk-dominated and peculiar counterparts (Figs.\,\ref{SSFRMassOffsets} and \ref{fig:MZmorph}) in agreement with an FMR scenario combined with the effect of dominant bulges lowering SFRs.

\begin{figure*}[!h]
\includegraphics[width=12cm,angle=270,clip=true]{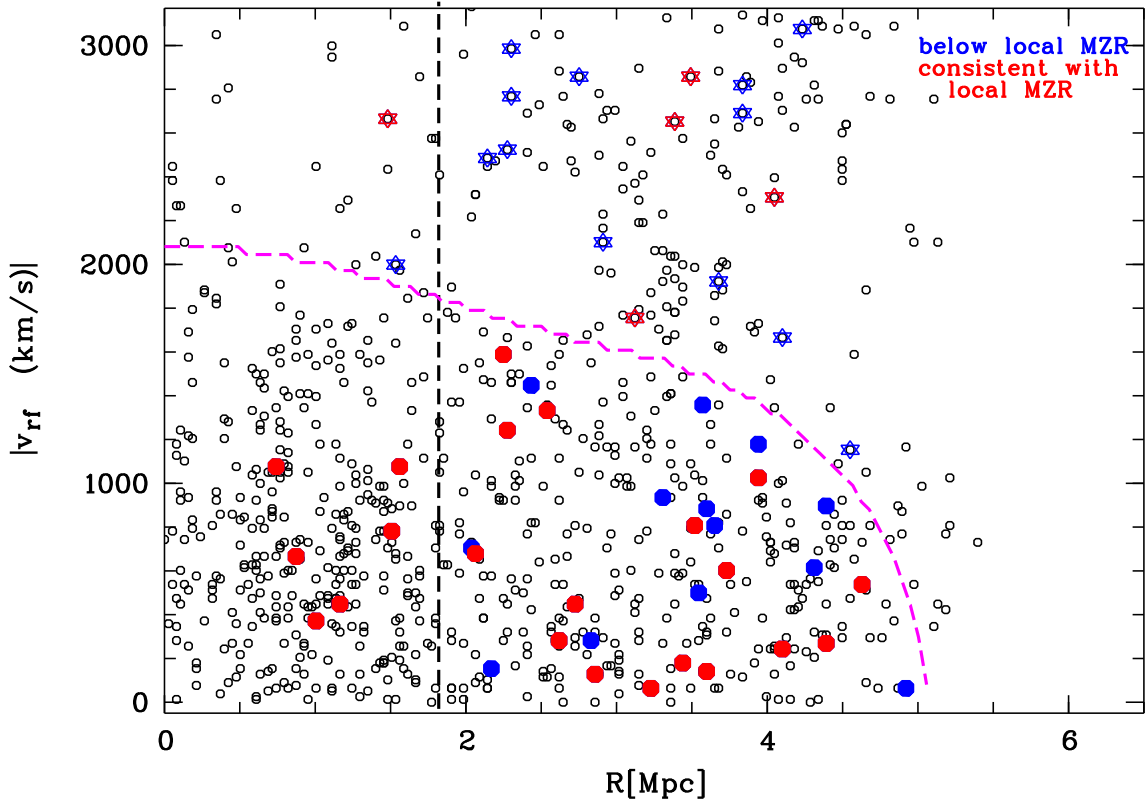}
\caption
{
\label{fig:PhaseSpace} 
\footnotesize 
Phase-space diagram based on \emph{all} MAC0416 spectroscopic members (black open circles). Our study concentrates on the objects for which we were able to measure metallicities  ($Zgals$ sample). Restricting ourselves to intermediate masses $9.2<\rm{log(M/M_{\odot})}<10.2$ we depict high metallicity objects (consistent with the local MZR)  as red symbols  and galaxies with  lower metallicities (below the local MZR) as blue symbols. The  dashed vertical line indicates $\rm{R}_{200} \sim 1.8\,\rm{Mpc}$ of the cluster (as reported in B16); the dashed magenta line indicates the caustic determined by B16.
 The diagram enables us to divide the cluster galaxies into infalling objects (above and to the right of the magenta line, star symbols) and accreted objects (filled circles).  
}
\end{figure*}

\begin{figure*}
\includegraphics[width=12cm,angle=270,clip=true]{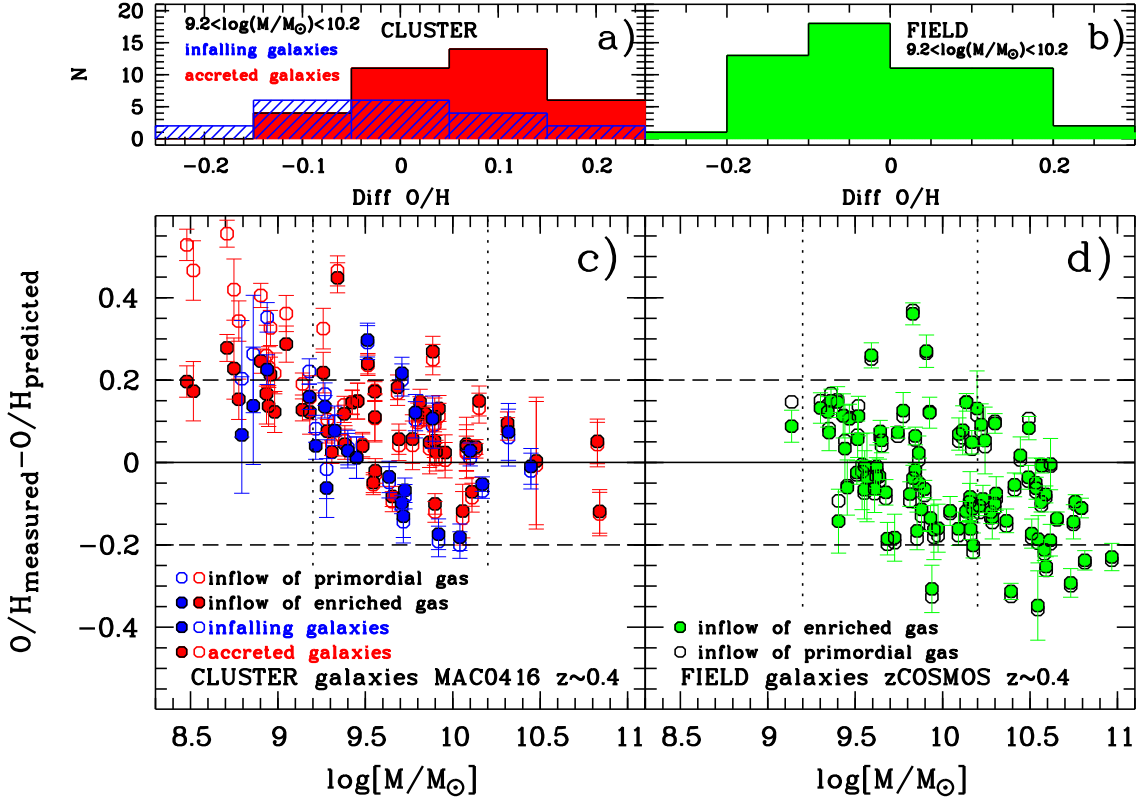}
\caption
{
\label{FMRDiff} 
\footnotesize 
Difference between the measured O/Hs of $z \sim 0.4$ $Zgals$ cluster galaxies (panel c) and zCOSMOS field objects (panel d) and the expected O/Hs from the formulations of Li13 for different infall metallicities $Z_{0}$ relative to the yield $y$, $Z_{0}/y=0.1$ (enriched inflow, filled circles) and $Z_{0}/y=0$ (pristine inflow, open circles). In addition, in  panel c)  the position of cluster galaxies in the phase-space diagram (see Fig.\,\ref{fig:PhaseSpace}) is indicated in red (blue) for accreted (infalling) galaxies. 
The upper histograms (panels a and b) for $\rm{M}_{\rm{med}}$ objects show  that in general  the infalling
cluster members (blue histogram) and field galaxies (panel b) have  lower metallicities, in quite good agreement with the FMR, unlike accreted galaxies (red histogram), which  in general have higher metallicities than predicted by the FMR. 
}
\end{figure*}

\begin{figure*}[!h]
\includegraphics[width=13cm,angle=270,clip=true]{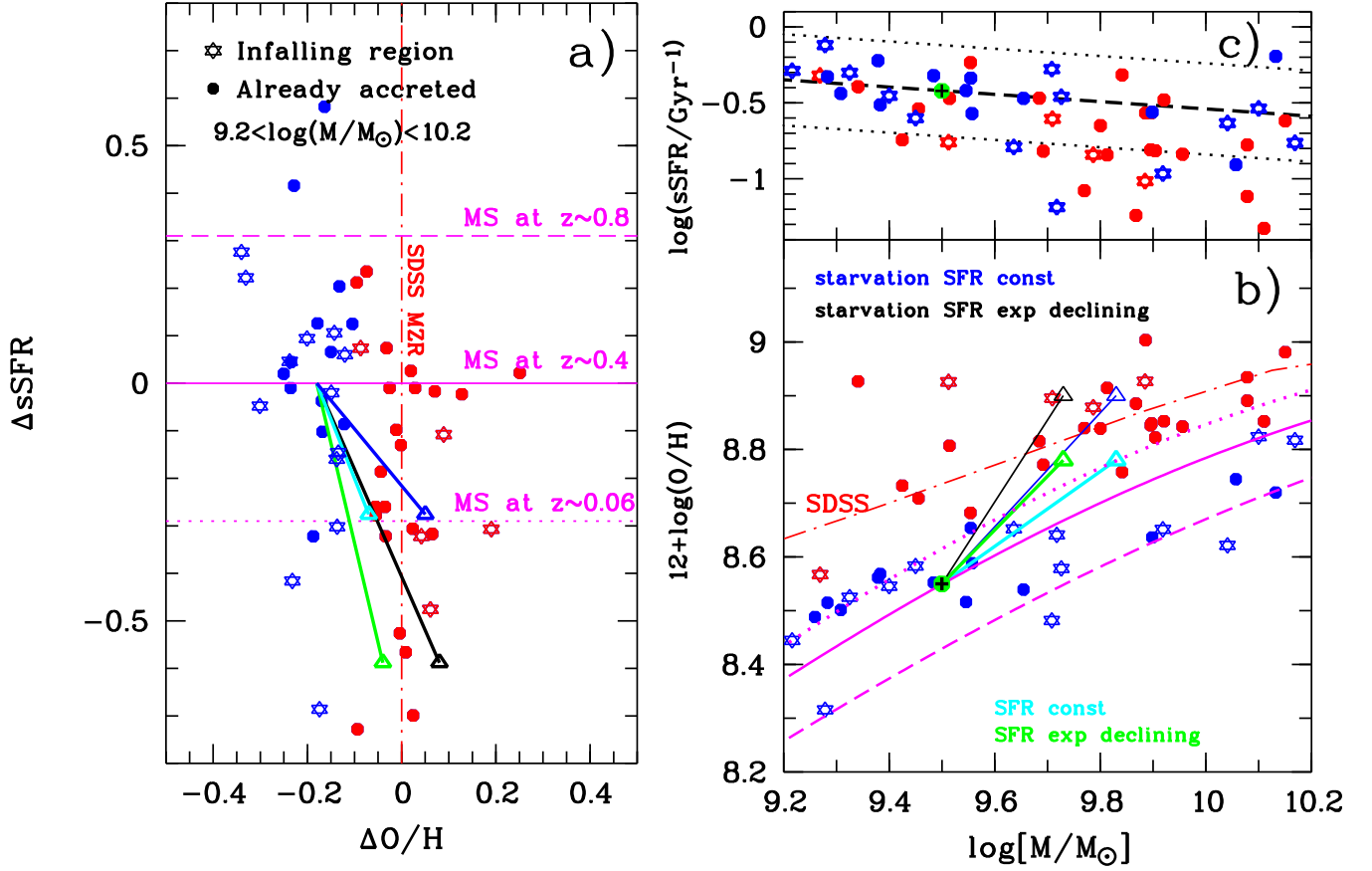}
\caption
{
\label{fig:DeltaOHDeltaSSFR} 
\footnotesize 
Panel a: Offsets $\Delta \rm{O/H}$ to the local MZR  vs. $\Delta \rm{sSFR}$ to the MS at $z \sim0.4$ for cluster galaxies with masses $9.2<\rm{log(M/M_{\odot})}<10.2$ ($\rm{M}_{\rm{med}}$). Filled circles (stars)  indicate accreted (infalling) galaxies based on their position in the phase-space diagram
(Fig.\,\ref{fig:PhaseSpace}). 
Galaxies with O/Hs consistent with the local MZR are shown in red, other objects in blue.
The MS mean values  at $z \sim 0.8$, $z \sim 0.4$, and $z \sim 0.06$, based on the equation of the evolution of the sSFR from \citet{peng10}, are indicated by dashed, solid, and dotted magenta lines.
The dash-dotted red line indicates the median local MZR  of our matched SDSS sample (see also Fig.\,\ref{MZRz07UK}).
Four tracks of a model galaxy start with $1.2\,\rm{M}_{\odot}$/yr and $\rm{log(M/M_{\odot})}=9.5$.
The model galaxy continues forming stars for 3\,Gyrs with a constant SFR (cyan track) or exponentially declining SFR with a $\tau =3$\,Gyrs (green track), and the corresponding tracks are shown for a strangulation scenario with constant SFR (blue track) and exponentially declining SFR (black track).  
Panel b: MZR for $\rm{M}_{\rm{med}}$ cluster galaxies with data symbols and  paths of model galaxies color-coded as in panel a. 
The magenta lines show the Li13 expectations of the Z(M,SFR) for galaxies on the MS at different redshifts, with dashed, solid, and dotted lines corresponding to different $z$ like the magenta lines in the panel a.
Panel c: The mass$-$sSFR relation for cluster galaxies color-coded in red for O/Hs consistent with the local MZR. Dashed and dotted lines show the median and  dispersion of the mass$-$sSFR relation according to \citet{peng10}.
The green filled circles with a black cross  in panels b and c indicate the starting point of the toy models described in Sect.\,\ref{sect:MZRpaths}: $\rm{log(\rm{M}_{\rm{start}}/M_{\odot})}=9.5$, $\rm{SFR}_{\rm{start}}=1.2 M_{\odot}/yr$, and $\rm{Z}_{\rm{start}}=8.55$. 
For the sake of visibility we do not plot error bars in Fig.\,\ref{fig:DeltaOHDeltaSSFR}; however, the error bars were already shown in the respective mass-sSFR (Fig.\,\ref{SSFRMassz07Env}) and MZR plot (Fig.\,\ref{MZRz07UK}) for galaxies of all stellar masses.
}
\end{figure*}

3. We find several indications for a strangulation scenario for galaxies with $\rm{log(M/M_{\odot})}<10.2$ infalling into  a cluster:

i) First, based on the location of galaxies in the projected phase-space diagram, a higher fraction of accreted galaxies (63\%) are metal-rich  compared to the fraction of 28\% of metal-rich infalling galaxies (Fig.\,\ref{fig:PhaseSpace}).

ii) Second, while the measured O/Hs of field and infalling cluster galaxies are in quite good agreement with the FMR, accreted galaxies have a distribution shifted to higher metallicities than predicted by the Li13 models (Fig.\,\ref{FMRDiff}).
This trend is intensified for low-mass ($\rm{log(M/M_{\odot})}<9.2$) accreted cluster galaxies, which have $2-3$ times higher metallicities than predicted by the models with inflowing pristine gas, indicating that a strangulation scenario in which gas inflow has stopped enhancing their metallicities is a plausible mechanism. 

iii) Third, the paths of model galaxies in the MZR and in the $\Delta \rm{sSFR} -\Delta \rm{O/H}$
diagrams show that strangulation is needed, in addition to a constant or exponentially declining SFR, to explain the higher metallicities of accreted cluster galaxies (Fig.\,\ref{fig:DeltaOHDeltaSSFR}).

The main conclusion is that a strangulation scenario is plausible to explain the metallicities of  infalling and accreted $\rm{log(M/M_{\odot})}<10.2$ cluster galaxies in MAC0416.
While the Z(M,SFR) measurements of most field and infalling cluster galaxies are in agreement with the FMR assuming a primordial gas inflow, we find higher values than from the FMR expected  metallicities  in accreted 
cluster galaxies, especially at lower masses, indicating that the inflow of pristine gas is being stopped in these galaxies as a result of cluster environmental effects.


\begin{acknowledgements}
We would like to thank the anonymous referee for providing constructive comments and help in improving the manuscript. This publication is supported by the Austrian Science Fund (FWF). IB, AM, PR, MN, and BS acknowledge support from PRIN-INAF 2014 1.05.01.94.02 (PI M. Nonino).
\end{acknowledgements}



%
%

\newcommand{\ra}[1]{\renewcommand{\arraystretch}{#1}}

\begin{table*}
\centering 
\ra{1.3}  
\begin{tabular}{@{}rrrrrcrrrrcrrrr@{}}\toprule 
& \multicolumn{4}{c}{$luminosity\ (R\ magnitude)$} & \phantom{abc} & \multicolumn{4}{c}{$stellar\ mass$} & 
     \phantom{abc} & \multicolumn{4}{c}{$color\ B-R$}\\ 

\cmidrule{2-5} \cmidrule{7-10} \cmidrule{12-15} 

$bins$  & $faint$ & $\rightarrow$ & $\rightarrow$ & $bright$ && $low$ & $\rightarrow$ & $\rightarrow$ &$high$ && $blue$ & $\rightarrow$ & $\rightarrow$ & $red$\\ 
\midrule 

$members$ & 21\%& 37\% & 32\% & 10\% && 15\% & 41\% & 29\% & 16\% && 13\% & 47\% & 25\% &15\%\\ 
$Zgals$ & 12\%& 39\% & 41\% & 8\% && 11\% & 39\% & 42\% & 8\% && 11\% & 55\% & 24\% &11\%\\ 
$fraction$ & 16\%& 30\% & 36\% & 22\% && 20\% & 27\% & 42\% & 14\% && 22\% & 33\% & 27\% &20\%\\

\bottomrule 
\end{tabular}
\caption{Demonstration of the coverage of our sample. First row: Percentage of spectroscopic members of MAC0416 representative of the blue cloud in the given bins in luminosity, mass, and color; 
Second row: selected $Zgals$ objects for this metallicity study in the given bins in luminosity, mass, and color;
Third row: the fractions of $Zgals$ objects in different luminosity/mass/color bins with respect to the parent sample of blue cluster members (numbers shown in Fig.\,\ref{Selec0416}).} 
\label{tab:percentages}
\end{table*}


\clearpage

\newpage

\begin{table*}
\ra{1.3}  
\caption{Observed and derived quantities for the 
 MAC0416 $z \sim 0.4$ cluster galaxies. EL fluxes are given in $10^{-17}\rm{ergs}\,\rm{s}^{-1}\rm{cm}^{-2}$.
The \Hb\, flux was corrected for stellar absorption as described in Sect.\,\ref{sec:elflux}, and all fluxes in the table were corrected for slit losses as described in Sect.\,\ref{sec:SFRs}. The stellar mass uncertainties were computed as described in Sect.\,\ref{sec:masses}.
}\label{MeasurementsMAC0416}
\begin{tabular}{cccccccc}
\hline\hline      
Id & z &    \Hb
  &   \OIIIa   &  \Ha   &  \NII  & O/H  & log(M/M$_{\sun}$) \\
\hline  
 20904 & 0.4007 &  9.75 $\pm$  0.97 &   6.48 $\pm$  0.68   & 35.82 $\pm$  1.71 & 10.23 $\pm$  1.71 &  8.85 $\pm$  0.03 &   9.92   $^{-0.19}_{+0.04}$ \\
 24665 & 0.3907 &  7.97 $\pm$  0.63 &  13.90 $\pm$  0.48   & 25.17 $\pm$  1.20 &  1.68 $\pm$  0.24 &  8.52 $\pm$  0.02 &   9.55   $^{-0.13}_{+0.04}$ \\
 25506 & 0.3844 &  5.04 $\pm$  0.63 &  15.15 $\pm$  0.64   & 17.57 $\pm$  0.97 &  1.93 $\pm$  0.32 &  8.51 $\pm$  0.03 &   9.18   $^{-0.03}_{+0.03}$ \\
 25980 & 0.3858 &  4.69 $\pm$  0.84 &   2.05 $\pm$  0.59   & 14.64 $\pm$  1.17 &  4.69 $\pm$  0.88 &  8.93 $\pm$  0.05 &   9.88   $^{-0.04}_{+0.03}$ \\
 27930 & 0.4048 &  6.82 $\pm$  0.53 &  12.01 $\pm$  0.42   & 18.96 $\pm$  0.84 &  3.37 $\pm$  0.84 &  8.65 $\pm$  0.04 &   9.92   $^{-0.02}_{+0.05}$ \\
 28862 & 0.3992 &  6.85 $\pm$  0.53 &  16.59 $\pm$  0.46   & 18.90 $\pm$  0.92 &  1.38 $\pm$  0.23 &  8.48 $\pm$  0.03 &   9.14   $^{-0.04}_{+0.04}$ \\
 31765 & 0.3931 &  2.50 $\pm$  0.22 &   6.94 $\pm$  0.35   &  8.33 $\pm$  0.69 &  0.52 $\pm$  0.17 &  8.44 $\pm$  0.05 &   8.98   $^{-0.04}_{+0.08}$ \\
 32731 & 0.4034 &  4.93 $\pm$  0.50 &  21.03 $\pm$  0.44   & 12.71 $\pm$  0.66 &  0.66 $\pm$  0.22 &  8.36 $\pm$  0.05 &   8.78   $^{-0.02}_{+0.11}$ \\
 33061 & 0.3839 &  1.63 $\pm$  0.25 &   6.32 $\pm$  0.33   &  4.58 $\pm$  0.44 &  0.22 $\pm$  0.22 &  8.36 $\pm$  0.14 &   8.79   $^{-0.11}_{+0.17}$ \\
 33592 & 0.3846 &  6.70 $\pm$  1.60 &   2.73 $\pm$  0.41   & 29.79 $\pm$  1.37 & 10.93 $\pm$  1.37 &  8.96 $\pm$  0.04 &  10.45   $^{-0.17}_{+0.01}$ \\
 43168 & 0.3899 &  2.07 $\pm$  0.26 &   4.27 $\pm$  0.21   &  5.33 $\pm$  0.64 &  0.43 $\pm$  0.21 &  8.52 $\pm$  0.07 &   8.93   $^{-0.09}_{+0.08}$ \\
 47541 & 0.3903 &  6.81 $\pm$  0.63 &  13.42 $\pm$  0.79   & 23.67 $\pm$  1.32 &  2.10 $\pm$  0.26 &  8.54 $\pm$  0.02 &   9.65   $^{-0.21}_{+0.04}$ \\
 48262 & 0.4040 &  1.98 $\pm$  0.40 &   0.66 $\pm$  0.11   &  7.25 $\pm$  0.44 &  1.76 $\pm$  0.22 &  8.93 $\pm$  0.04 &   9.51   $^{-0.24}_{+0.03}$ \\
 54317 & 0.3928 & 11.92 $\pm$  1.20 &  25.00 $\pm$  0.63   & 33.13 $\pm$  1.25 &  3.44 $\pm$  0.63 &  8.55 $\pm$  0.03 &   9.48   $^{-0.03}_{+0.05}$ \\
 55515 & 0.3934 &  7.98 $\pm$  1.49 &   2.22 $\pm$  0.28   & 34.08 $\pm$  2.77 & 10.25 $\pm$  1.11 &  8.98 $\pm$  0.04 &  10.15   $^{-0.02}_{+0.21}$ \\
 60453 & 0.3875 & 16.85 $\pm$  0.81 &  15.26 $\pm$  0.66   & 51.70 $\pm$  1.64 & 15.59 $\pm$  1.64 &  8.82 $\pm$  0.02 &  10.17   $^{-0.02}_{+0.05}$ \\
 61634 & 0.3843 & 11.06 $\pm$  0.65 &  21.31 $\pm$  0.81   & 35.30 $\pm$  0.81 &  4.07 $\pm$  0.81 &  8.58 $\pm$  0.03 &   9.73   $^{-0.03}_{+0.04}$ \\
 62697 & 0.3814 &  5.76 $\pm$  2.12 &   1.77 $\pm$  0.55   & 26.82 $\pm$  1.33 &  9.98 $\pm$  1.11 &  9.00 $\pm$  0.07 &  10.32   $^{-0.02}_{+0.12}$ \\
 64825 & 0.4026 &  7.52 $\pm$  0.92 &  14.06 $\pm$  0.66   & 20.76 $\pm$  1.31 &  2.50 $\pm$  0.39 &  8.59 $\pm$  0.03 &   9.56   $^{-0.12}_{+0.04}$ \\
 65028 & 0.3844 &  5.90 $\pm$  0.47 &  11.22 $\pm$  0.64   & 18.21 $\pm$  0.64 &  1.91 $\pm$  0.32 &  8.57 $\pm$  0.03 &   9.27   $^{-0.23}_{+0.00}$ \\
 68953 & 0.4092 &  2.54 $\pm$  0.38 &   1.35 $\pm$  0.25   &  9.82 $\pm$  0.85 &  2.71 $\pm$  0.51 &  8.88 $\pm$  0.04 &   9.79   $^{-0.13}_{+0.04}$ \\
 69388 & 0.3836 &  3.55 $\pm$  0.38 &   6.99 $\pm$  0.33   & 12.03 $\pm$  0.98 &  1.46 $\pm$  0.49 &  8.58 $\pm$  0.05 &   9.45   $^{-0.33}_{+0.01}$ \\
 69442 & 0.3951 & 26.39 $\pm$  1.12 &  23.83 $\pm$  0.78   & 75.12 $\pm$  2.59 & 14.76 $\pm$  1.30 &  8.76 $\pm$  0.02 &   9.84   $^{-0.04}_{+0.03}$ \\
 70675 & 0.4108 &  3.84 $\pm$  0.41 &  11.52 $\pm$  0.35   & 12.74 $\pm$  0.70 &  0.87 $\pm$  0.17 &  8.44 $\pm$  0.03 &   9.22   $^{-0.02}_{+0.05}$ \\
 72228 & 0.3894 &  2.12 $\pm$  0.21 &  12.15 $\pm$  0.37   &  6.99 $\pm$  0.28 &  0.37 $\pm$  0.09 &  8.32 $\pm$  0.04 &   8.48   $^{-0.05}_{+0.08}$ \\
 73404 & 0.3954 &  5.97 $\pm$  0.50 &   4.41 $\pm$  0.37   & 19.29 $\pm$  0.92 &  4.41 $\pm$  0.55 &  8.81 $\pm$  0.02 &   9.51   $^{-0.03}_{+0.06}$ \\
 74065 & 0.4002 &  5.91 $\pm$  0.54 &  14.94 $\pm$  0.47   & 16.11 $\pm$  0.70 &  1.40 $\pm$  0.23 &  8.50 $\pm$  0.03 &   9.31   $^{-0.16}_{+0.11}$ \\
 76447 & 0.4084 & 14.16 $\pm$  1.31 &  24.84 $\pm$  0.87   & 44.22 $\pm$  1.31 &  6.32 $\pm$  1.31 &  8.62 $\pm$  0.03 &  10.04   $^{-0.11}_{+0.02}$ \\
 78137 & 0.3932 & 14.01 $\pm$  4.88 &   9.57 $\pm$  0.93   & 63.05 $\pm$  4.67 & 25.69 $\pm$  2.34 &  8.90 $\pm$  0.05 &  10.84   $^{-0.04}_{+0.02}$ \\
 80425 & 0.4090 &  5.87 $\pm$  1.11 &   7.22 $\pm$  0.60   & 14.45 $\pm$  0.90 &  1.81 $\pm$  0.30 &  8.65 $\pm$  0.04 &   9.64   $^{-0.05}_{+0.03}$ \\
 80427 & 0.4088 & 10.58 $\pm$  1.00 &   8.40 $\pm$  0.51   & 40.89 $\pm$  1.01 & 11.33 $\pm$  0.81 &  8.82 $\pm$  0.02 &  10.10   $^{-0.04}_{+0.05}$ \\
 82461 & 0.3865 &  4.81 $\pm$  0.43 &  11.19 $\pm$  0.60   & 14.77 $\pm$  1.19 &  1.07 $\pm$  0.24 &  8.49 $\pm$  0.04 &   8.94   $^{-0.03}_{+0.09}$ \\
 84626 & 0.3967 &  3.38 $\pm$  0.40 &  10.94 $\pm$  0.50   & 12.26 $\pm$  0.83 &  1.08 $\pm$  0.17 &  8.47 $\pm$  0.03 &   8.90   $^{-0.08}_{+0.03}$ \\
 85922 & 0.3935 &  5.60 $\pm$  0.61 &  18.92 $\pm$  0.66   & 17.60 $\pm$  0.88 &  1.10 $\pm$  0.22 &  8.42 $\pm$  0.03 &   8.71   $^{-0.09}_{+0.03}$ \\
 87340 & 0.3939 & 12.37 $\pm$  2.96 &   2.52 $\pm$  0.76   & 49.73 $\pm$  1.77 & 22.72 $\pm$  1.26 &  9.08 $\pm$  0.05 &  10.83   $^{-0.03}_{+0.04}$ \\
 92461 & 0.3917 &  4.68 $\pm$  0.71 &   3.12 $\pm$  0.32   & 13.56 $\pm$  0.65 &  3.87 $\pm$  0.65 &  8.85 $\pm$  0.04 &  10.11   $^{-0.02}_{+0.05}$ \\
 92623 & 0.3922 &  3.56 $\pm$  0.76 &   6.24 $\pm$  0.56   & 12.47 $\pm$  0.89 &  3.34 $\pm$  0.67 &  8.71 $\pm$  0.04 &   9.46   $^{-0.06}_{+0.02}$ \\
 92946 & 0.3998 &  6.54 $\pm$  1.49 &   6.61 $\pm$  0.43   & 20.99 $\pm$  0.86 &  8.63 $\pm$  1.44 &  8.84 $\pm$  0.04 &   9.90   $^{-0.01}_{+0.06}$ \\
 93590 & 0.3940 &  4.11 $\pm$  0.43 &   1.42 $\pm$  0.32   & 14.18 $\pm$  0.63 &  3.78 $\pm$  0.63 &  8.93 $\pm$  0.04 &  10.08   $^{-0.17}_{+0.02}$ \\
 95694 & 0.4071 &  8.17 $\pm$  0.48 &  14.60 $\pm$  0.40   & 22.00 $\pm$  0.80 &  1.60 $\pm$  0.40 &  8.53 $\pm$  0.04 &   9.32   $^{-0.00}_{+0.43}$ \\
\hline
\end{tabular}
\end{table*}

\newpage

\begin{table*}
\ra{1.3}  
\caption{(Table 2 continued)\label{MeasurementsMAC0416B}}
\begin{tabular}{cccccccc}
\hline\hline      
Id & z &    \Hb
  &   \OIIIa    &   \Ha    &  \NII   & O/H  & log(M/M$_{\sun}$) \\
\hline      
105374 & 0.3998 &   8.54 $\pm$ 0.82 &  2.36 $\pm$  0.34 & 39.45 $\pm$  1.01 & 11.13 $\pm$  1.01     &  8.97 $\pm$  0.03 &  10.31   $^{-0.27}_{+0.00}$ \\
106119 & 0.4058 &   8.09 $\pm$ 0.83 & 23.92 $\pm$  1.01 & 25.27 $\pm$  1.01 &  0.67 $\pm$  0.34     &  8.32 $\pm$  0.07 &   9.28   $^{-0.03}_{+0.04}$ \\
107694 & 0.3975 &   1.54 $\pm$ 0.24 &  4.73 $\pm$  0.19 &  5.98 $\pm$  0.58 &  0.39 $\pm$  0.19     &  8.43 $\pm$  0.07 &   8.75   $^{-0.04}_{+0.12}$ \\
108060 & 0.3975 &   6.44 $\pm$ 1.71 &  5.23 $\pm$  0.58 & 25.19 $\pm$  1.94 &  8.52 $\pm$  1.16     &  8.85 $\pm$  0.05 &   9.90   $^{-0.05}_{+0.03}$ \\
110267 & 0.3964 &   3.68 $\pm$ 1.43 &  1.84 $\pm$  0.18 & 16.92 $\pm$  1.47 &  4.78 $\pm$  1.10     &  8.89 $\pm$  0.07 &  10.08   $^{-0.08}_{+0.03}$ \\
110762 & 0.3851 &   2.50 $\pm$ 0.47 &  6.08 $\pm$  0.54 &  7.51 $\pm$  0.72 &  0.36 $\pm$  0.36     &  8.42 $\pm$  0.14 &   8.86   $^{-0.04}_{+0.09}$ \\
113035 & 0.4022 &   4.76 $\pm$ 2.16 &  0.95 $\pm$  0.95 & 17.87 $\pm$  1.08 &  4.60 $\pm$  0.81     &  9.01 $\pm$  0.15 &  10.48   $^{-0.03}_{+0.04}$ \\
115312 & 0.3892 &   6.09 $\pm$ 0.93 &  3.42 $\pm$  0.43 & 20.95 $\pm$  1.50 &  4.70 $\pm$  1.07     &  8.84 $\pm$  0.04 &   9.96   $^{-0.04}_{+0.04}$ \\
116392 & 0.3959 &   3.57 $\pm$ 0.34 &  3.58 $\pm$  0.38 & 10.36 $\pm$  0.47 &  1.88 $\pm$  0.28     &  8.73 $\pm$  0.03 &   9.42   $^{-0.13}_{+0.05}$ \\
118060 & 0.3836 &   3.62 $\pm$ 0.56 &  1.66 $\pm$  0.25 & 14.79 $\pm$  1.02 &  3.95 $\pm$  0.64     &  8.90 $\pm$  0.04 &   9.71   $^{-0.03}_{+0.19}$ \\
119315 & 0.4007 &  14.41 $\pm$ 1.26 & 41.30 $\pm$  1.75 & 46.90 $\pm$  3.50 &  4.20 $\pm$  1.40     &  8.49 $\pm$  0.05 &   9.26   $^{-0.02}_{+0.05}$ \\
119393 & 0.4001 &   2.26 $\pm$ 0.42 &  1.10 $\pm$  0.16 &  9.28 $\pm$  0.63 &  3.30 $\pm$  0.31     &  8.93 $\pm$  0.04 &   9.34   $^{-0.04}_{+0.02}$ \\
120959 & 0.3822 &  10.39 $\pm$ 0.60 & 30.12 $\pm$  0.53 & 38.52 $\pm$  0.88 &  3.33 $\pm$  0.35     &  8.48 $\pm$  0.02 &   9.71   $^{-0.42}_{+0.01}$ \\
121798 & 0.3962 &  26.35 $\pm$ 1.91 & 31.40 $\pm$  0.86 &102.76 $\pm$  2.85 & 20.27 $\pm$  2.00     &  8.72 $\pm$  0.02 &  10.13   $^{-0.02}_{+0.24}$ \\
122137 & 0.3969 &   5.20 $\pm$ 0.59 & 14.43 $\pm$  0.72 & 14.43 $\pm$  0.96 &  1.32 $\pm$  0.36     &  8.50 $\pm$  0.04 &   8.96   $^{-0.02}_{+0.11}$ \\
122217 & 0.4013 &   5.37 $\pm$ 0.78 &  8.32 $\pm$  0.54 & 15.56 $\pm$  1.07 &  1.34 $\pm$  0.27     &  8.57 $\pm$  0.04 &   9.38   $^{-0.03}_{+0.08}$ \\
122730 & 0.3909 &   3.99 $\pm$ 0.90 &  2.28 $\pm$  0.25 & 11.31 $\pm$  0.84 &  2.53 $\pm$  0.84     &  8.84 $\pm$  0.06 &   9.77   $^{-0.04}_{+0.04}$ \\
123859 & 0.3956 &   1.62 $\pm$ 0.60 &  3.24 $\pm$  0.27 &  9.18 $\pm$  0.54 &  1.89 $\pm$  0.27     &  8.65 $\pm$  0.06 &   9.55   $^{-0.05}_{+0.03}$ \\
126188 & 0.3943 &   2.72 $\pm$ 0.36 & 13.11 $\pm$  0.34 &  7.65 $\pm$  0.25 &  0.34 $\pm$  0.17     &  8.32 $\pm$  0.07 &   8.52   $^{-0.06}_{+0.03}$ \\
127932 & 0.3949 &   5.08 $\pm$ 0.62 &  5.08 $\pm$  0.30 & 15.23 $\pm$  0.81 &  3.65 $\pm$  0.81     &  8.77 $\pm$  0.04 &   9.69   $^{-0.05}_{+0.02}$ \\
128401 & 0.3904 &   1.99 $\pm$ 0.74 &  1.16 $\pm$  0.17 &  6.80 $\pm$  0.66 &  2.16 $\pm$  0.33     &  8.89 $\pm$  0.06 &   9.87   $^{-0.05}_{+0.06}$ \\
128987 & 0.3957 &   4.48 $\pm$ 0.59 &  4.08 $\pm$  0.31 & 16.47 $\pm$  0.78 &  5.18 $\pm$  0.63     &  8.82 $\pm$  0.03 &   9.90   $^{-0.02}_{+0.06}$ \\
130603 & 0.3997 &   3.18 $\pm$ 0.32 &  8.55 $\pm$  0.40 & 11.73 $\pm$  0.60 &  1.19 $\pm$  0.20     &  8.51 $\pm$  0.03 &   9.28   $^{-0.09}_{+0.04}$ \\
131120 & 0.3954 &   1.33 $\pm$ 0.20 &  2.31 $\pm$  0.18 &  4.70 $\pm$  0.44 &  0.53 $\pm$  0.18     &  8.59 $\pm$  0.05 &   9.18   $^{-0.15}_{+0.12}$ \\
131278 & 0.3955 &   5.03 $\pm$ 0.66 &  2.44 $\pm$  0.27 & 16.27 $\pm$  0.95 &  5.29 $\pm$  0.68     &  8.91 $\pm$  0.03 &   9.81   $^{-0.11}_{+0.18}$ \\
132372 & 0.4022 &   1.47 $\pm$ 0.41 &  2.14 $\pm$  0.15 &  5.05 $\pm$  0.31 &  0.69 $\pm$  0.15     &  8.64 $\pm$  0.05 &   9.72   $^{-0.15}_{+0.04}$ \\
133372 & 0.3834 &   3.38 $\pm$ 0.23 &  5.89 $\pm$  0.29 & 12.53 $\pm$  0.59 &  1.03 $\pm$  0.29     &  8.55 $\pm$  0.04 &   9.40   $^{-0.16}_{+0.06}$ \\
136237 & 0.3950 &   2.50 $\pm$ 0.93 &  2.71 $\pm$  0.21 & 11.69 $\pm$  1.25 &  2.50 $\pm$  0.63     &  8.74 $\pm$  0.07 &  10.06   $^{-0.27}_{+0.02}$ \\
140223 & 0.3967 &   3.99 $\pm$ 0.37 &  2.99 $\pm$  0.33 & 15.94 $\pm$  0.66 &  4.65 $\pm$  0.66     &  8.84 $\pm$  0.03 &   9.80   $^{-0.03}_{+0.17}$ \\
141465 & 0.3944 &   2.72 $\pm$ 0.31 &  4.29 $\pm$  0.19 &  8.96 $\pm$  0.47 &  1.31 $\pm$  0.37     &  8.64 $\pm$  0.04 &   9.05   $^{-0.11}_{+0.15}$ \\
141596 & 0.3978 &   3.74 $\pm$ 0.43 &  3.15 $\pm$  0.26 & 15.99 $\pm$  0.85 &  4.42 $\pm$  0.68     &  8.82 $\pm$  0.03 &   9.68   $^{-0.41}_{+0.00}$ \\
144037 & 0.4016 &   8.01 $\pm$ 0.68 & 13.21 $\pm$  0.88 & 28.61 $\pm$  2.20 &  5.94 $\pm$  0.66     &  8.68 $\pm$  0.02 &   9.55   $^{-0.05}_{+0.02}$ \\
144528 & 0.4021 &   6.12 $\pm$ 0.51 & 14.62 $\pm$  0.28 & 21.09 $\pm$  1.41 &  2.67 $\pm$  0.28     &  8.56 $\pm$  0.02 &   9.38   $^{-0.04}_{+0.05}$ \\
144606 & 0.3993 &   4.05 $\pm$ 0.45 &  1.21 $\pm$  0.26 & 18.18 $\pm$  0.87 &  6.93 $\pm$  0.69     &  9.00 $\pm$  0.04 &   9.89   $^{-0.02}_{+0.17}$ \\
148186 & 0.3981 &   2.35 $\pm$ 0.24 &  7.15 $\pm$  0.39 &  6.95 $\pm$  0.39 &  0.58 $\pm$  0.19     &  8.47 $\pm$  0.05 &   8.95   $^{-0.05}_{+0.14}$ \\
148883 & 0.3970 &   5.40 $\pm$ 0.47 &  9.16 $\pm$  0.35 & 22.46 $\pm$  0.69 &  3.46 $\pm$  0.52     &  8.64 $\pm$  0.03 &   9.90   $^{-0.36}_{+0.00}$ \\
\hline
\end{tabular}
\end{table*}


\end{document}